\documentclass[12pt]{article}
\usepackage{amsmath,amssymb,bm,epsfig,color,graphicx,cite,braket,mathrsfs,slashed}
\renewcommand{\thefootnote}{\fnsymbol{footnote}}

\begin{document}

\title{
\begin{flushright}
\begin{minipage}{0.2\linewidth}
\normalsize
\end{minipage}
\end{flushright}
{\Large \bf 
Precise Capture Rates of Cosmic Neutrinos and Their Implications on Cosmology
\\*[20pt]}}

\author{
Kensuke Akita$^{1}$\footnote{
E-mail address: kensuke@th.phys.titech.ac.jp}\ ,\  
Saul Hurwitz$^{1}$\footnote{
E-mail address: hurwitz.s.aa@m.titech.ac.jp}\ and\
Masahide Yamaguchi$^{1}$\footnote{
E-mail address: gucci@phys.titech.ac.jp}\\*[20pt]
$^1${\it \normalsize
Department of Physics, Tokyo Institute of Technology,
Tokyo 152-8551, Japan} \\*[50pt]
}

\date{
\centerline{\small \bf Abstract}
\begin{minipage}{0.9\linewidth}
\medskip \medskip \small 
We explore the potential of measurements of cosmological
effects, such as neutrino spectral distortions from the neutrino
decoupling and neutrino clustering in our Galaxy, via cosmic neutrino
capture on tritium. We compute the precise capture rates of each neutrino
species including such cosmological effects to probe them.
These precise estimates of capture rates are also important in that the
would-be deviation of the estimated capture rate could suggest new
neutrino physics and/or a non-standard evolution of the universe. 
In addition, we discuss the precise differences between the capture rates
of Dirac and Majorana neutrinos for each species, the required energy
resolutions to detect each neutrino species and the method of reconstruction
of the spectrum of cosmic neutrinos via the spectrum of emitted electrons, with emphasis
on the PTOLEMY experiment. 
\end{minipage}
}

\maketitle{}
\thispagestyle{empty}
\clearpage
\tableofcontents
\clearpage

\renewcommand{\thefootnote}{\arabic{footnote}}
\setcounter{footnote}{0}

\section{Introduction}
\label{sec:1}

Garnering massive predictive success, the standard big bang theory is adept at explaining a plethora of cosmological and astrophysical phenomena. If the model is correct, at around 1 second after the formation of the universe, neutrinos would have decoupled from the interacting particles in the universe. Analogous to photons that make up the CMB, these decoupled neutrinos are expected to have been free streaming until today, and can provide information from
the time they decoupled at \rm{MeV} scale temperature. If these relic neutrinos, collectively called the cosmic neutrino background (C$\nu$B), are detected, then we can probe information about our universe at much earlier times than we are currently able as well as uncovering the properties of neutrinos themselves. The existence of these neutrinos is strongly supported by indirect evidence such as the observational data of the primordial abundances of light elements from Big Bang Nucleosynthesis (BBN), the anisotropies of the Cosmic Microwave Background (CMB) and the distribution of Large Scale Structure (LSS) in the universe.
In particular, observations from the Planck satellite impose the severe constraint on the effective number of relativistic species, $N_{\rm eff}$, and the sum of the neutrino masses at $95\%$ CL as \cite{Aghanim:2018eyx}
\begin{align}
&N_{\rm eff} \equiv \frac{8}{7}\left(\frac{11}{4} \right)^{4/3}\left[\frac{\rho_r}{\rho_{\gamma}}-1 \right] = 2.99^{+0.34}_{-0.33} \ \ \ \  {\rm and}\ \ \ \  \sum m_{\nu}< 0.12\ {\rm eV},
\end{align}
where $\rho_r$ and $\rho_{\gamma}$ are the energy densities of radiation
and photons, respectively.  The theoretical value for this parameter
in the Standard Model (SM) is $N_{\rm eff}=3.044$ \cite{Mangano:2005cc,
deSalas:2016ztq, Bennett:2019ewm, Escudero:2020dfa}, (for the most precise calculations see \cite{Akita:2020szl,Froustey:2020mcq}), which is consistent
with the above constraints.  The next generation of cosmological
observations are expected to determine $N_{\rm eff}$ with 1\% precision
in the near future \cite{Abazajian:2013oma, DiValentino:2016foa,
Hanany:2019lle, Sehgal:2019ewc, Abazajian:2019eic}.

Unfortunately, the C$\nu$B has not yet been observed in a direct
way. Owing to the expansion of the universe, these neutrinos would have
lost most of their momenta, and will have very low energies, and so are
very difficult to detect.  Nevertheless, it is conceivable that the relic 
neutrinos would one day be detected directly. Such a direct probe of
the C$\nu$B would not only confirm that these neutrinos still exist in
the present universe but also complement our knowledge of cosmology and
neutrinos. Through the direct observation of cosmic neutrinos, we would
distinguish whether the origin of $N_{\rm eff}$ lies in neutrino species or
exotic relativistic species and/or from thermal or non-thermal neutrinos.
In other words, it would test many cosmological models wherein cosmic neutrinos decayed during some period of the universe in the majoron models \cite{Chikashige:1980ui, Gelmini:1980re, Schechter:1981cv}. It would also test if they were produced less in very low reheating scenarios
\cite{Kawasaki:1999na, Kawasaki:2000en, Hannestad:2004px, Ichikawa:2005vw, deSalas:2015glj, Hasegawa:2019jsa};
if their spectra and energy densities were modified by the decay of a heavy particle into neutrinos (see e.g. \cite{McKeen:2018xyz, Chacko:2018uke, Escudero:2019gzq, Jaeckel:2020oet});
or if some dark radiation contributes to $N_{\rm eff}$.

The most promising method of a direct detection of the C$\nu$B is via
neutrino capture on $\beta$-decaying nuclei (NCB)
\cite{Weinberg:1962zza, Cocco:2007za}, in particular on tritium
\cite{Lazauskas:2007da, Blennow:2008fh, Li:2010sn, Long:2014zva,
Arteaga:2017zxg, Roulet:2018fyh}, through the inverse $\beta$-decay process, $\nu + n
\rightarrow p + e^-$. Since there is no threshold energy in this
process, the inverse $\beta$-decay processes for neutrinos with
arbitrary energies are always allowed. The challenges of the NCB method
include the availability of $\beta$-decaying nuclei with long lifetimes
and the need for extremely high precision in measuring the outgoing
electron energy. As a $\beta$-decaying nucleus, tritium is an
appropriate candidate due to its availability, high neutrino capture
cross section, low Q-value and long lifetime with a half-life of
$t_{1/2} = 12.32$ years. In this method using a tritium target, 
provided that an extremely good energy resolution can be obtained, the
signature of the capture of one neutrino species $\nu_i$ with energy $E_{\nu_i}$ is a peak in the electron energy
spectrum at an energy of $(m_{\rm lightest} + E_{\nu_i})$ above the
$\beta$ decay endpoint\footnote{If the captured neutrinos with mass $m_{\nu_i}$ are non-relativistic, $m_{\rm lightest}+E_{\nu_i}\simeq m_{\rm lightest} + m_{\nu_i}$.} 
, where $m_{\rm lightest}$ is the lightest mass species
of neutrinos.
A planned project, formerly known as PTOLEMY (PonTecorvo Tritium Observatory for
Light, Early-Universe, Massive-Neutrino Yield) has begun 
developing an innovative technology that can improve the energy
resolution, envisaging the use of a $100$ gram tritium target
\cite{Baracchini:2018wwj, Betti:2018bjv, Betti:2019ouf}.
 
The theoretical calculation of the rate at which this experiment
will detect cosmic neutrinos is vital for extracting the various
properties of neutrinos from the experimental data, including their
actual masses, whether they are Dirac or Majorana fermions, and the
number of species. 
In addition, this capture rate is also important for probing and constraining cosmological effects on neutrinos such as gravitational clustering of neutrinos by our Galaxy, the nonequilibrium corrections to the neutrino spectra in the early universe, and lepton asymmetry in the neutrino sector.

In this paper, we explore the potential of measurements and constraints on such cosmological effects via cosmic neutrino capture on tritium in more detail.
In particular, we give the precise estimate of
cosmic neutrino capture rates on tritium. For this purpose, we go beyond
the leading order calculations of these capture rates, which have been
done before, and give the sub-leading order corrections including
neutrino spectral distortions from their decoupling and the gravitational
clustering of neutrinos. In fact, the possible deviation of the would-be observed capture rate from the estimated precise one could allow us to distinguish more accurately new neutrino physics
and/or non-standard evolution of the universe from the standard cosmology.
In addition, we comprehensively discuss the precise differences between the capture rates of Dirac and Majorana neutrinos and the required energy resolutions of the detector for each neutrino species of mass-eigenstate since one signal from the C$\nu$B would come from one mass-eigenstate of neutrino.
In this discussion, we consider both the normal and inverted mass hierarchies of neutrinos.
We also consider the reconstruction method of the neutrino spectrum from the observed spectrum of the emitted electrons.

This paper is organized as follows. In the next section, we discuss
the properties of the cosmic neutrino background including cosmological
effects such as the neutrino spectral distortion in the neutrino
decoupling and gravitational clustering of relic neutrinos. 
In particular, we estimate the precise number densities of neutrinos in the current universe,
including such cosmological effects.
In section \ref{sec:3}, the formulae of the neutrino capture rate including
cosmological effects are given. In section \ref{sec:4}, the precise
estimate of the neutrino capture rate as well as the required energy
resolution for the actual observation is given. The reconstruction
method of the neutrino spectrum from the would-be observed spectrum of
electrons is also discussed. The final section is devoted to summary and
discussion. In the appendix, the exact neutrino capture rate at tree level on the tritium target as well as the kinematics of tritium beta decay and inverse tritium beta decay for cosmic neutrinos are discussed.


\section{Cosmology of the C$\nu$B}
\label{sec:2}

In this section, we consider the cosmology of relic neutrinos. First, we review the history of the C$\nu$B in the instantaneous decoupling limit. 
Then, we discuss the sub-leading cosmological effects such as the neutrino spectral distortions in the neutrino decoupling and gravitational clustering of relic neutrinos.
In particular, we calculate the precise number density of cosmic neutrinos in the present universe, including the neutrino spectral distortions in the decoupling and gravitational clustering.

\subsection{Neutrino cosmology in the instantaneous decoupling limit}
\label{sec:2.1}

First, we consider the production and decoupling processes of neutrinos in the early universe. After that, we discuss the properties of neutrinos in the current universe.
Finally, with emphasis on the distinction between Dirac and Majorana neutrinos, we discuss which spin states of neutrinos are populated in the current universe. Here we basically follow the arguments given in \cite{Long:2014zva}.

It is predicted that at early times in the universe, when temperatures were much higher, that left-handed neutrinos and right-handed anti-neutrinos reacted and were produced constantly in thermal equilibrium with themselves and charged leptons via the weak interaction. In this epoch, left-handed neutrinos and right-handed anti-neutrinos were in thermal equilibrium and the shape of the spectrum for these massive neutrinos is the Fermi-Dirac distribution,
\begin{align}
f_{\rm FD}(\bm{p},T)=\frac{1}{e^{E/T}+1},\ \ \ \ E=\sqrt{\bm{p}^2+m_{\nu}^2},
\end{align}
where $T$ is the temperature of the thermal plasma. Here and hereafter, we assume lepton asymmetry is negligibly small. This assumption is reasonable since neutrino oscillations leading to flavor equilibrium before BBN impose
a stringent constraint on this asymmetry \cite{Dolgov:2002ab, Wong:2002fa, Abazajian:2002qx, Mangano:2011ip, Castorina:2012md, Oldengott:2017tzj}. In addition, the standard baryogenesis scenarios via the sphaleron process in leptogenesis models predict that the lepton asymmetry is of the order of the current baryon asymmetry, $n_B/n_{\gamma}\sim 10^{-10}$, which is much smaller than the above constraint.

Since neutrino masses are much smaller than the temperature in the early universe, the number density of one flavor (or mass) eigenstate for left-handed neutrinos is given by
\begin{align}
n_{\nu}(T)=\frac{3\zeta(3)}{4\pi^2}T^3.
\label{ND}
\end{align}

When the temperature of the plasma decreased due to the expansion of the universe, neutrinos did not interact with other particles and deviated from thermal equilibrium. This decoupling happened when the mean free travel time of neutrinos would have been comparable to the Hubble time. We can estimate a ball-park figure for this decoupling time using the mean free time, $\tau\approx\frac{1}{G_F^2T^5}$, and the Hubble time, $t_H\equiv H^{-1}\approx\frac{M_P}{T^2}$, where $G_F$ is the Fermi coupling constant and $M_P$ is the reduced Planck mass. Then, we find the approximate decoupling temperature to be around $2\ {\rm MeV}$, which corresponds to the period when the universe was around 1 second old.

After the decoupling, since the time dependence of the distribution function of free particles is determined only by the redshift of momenta, the neutrino distribution function is given by
\begin{align}
f_{\nu}(\bm{p},t)=\frac{1}{e^{|\bm{p}|/T_{\nu}(t)}+1},
\end{align}
where $T_{\nu}$ is the effective neutrino temperature,
\begin{align}
T_{\nu}(t)=\frac{a(t_{\rm dec})}{a(t)}T_{\rm dec}.
\end{align}
Here $a(t)$ is the scale factor of the universe, $t_{\rm dec}$ is the time of the decoupling, and $T_{\rm dec} \sim 2\ {\rm MeV}$ is the decoupling temperature of neutrinos. Here, we have assumed that the neutrinos were decoupled instantaneously without any momentum dependence.
After the decoupling of neutrinos, the photon temperature also decreased. When the photon temperature dropped below the electron mass, electrons and positrons annihilated into photons, injecting energy into this component. Due to this process, the photon temperature below the electron mass satisfies the following relation with the effective neutrino temperature, using entropy conservation,
\begin{align}
\frac{T_{\gamma}(t)}{T_{\nu}(t)}=\left(\frac{g_{\ast}(t_{\rm dec})}{g_{\ast}(t)} \right)^{1/3}=\left(\frac{11}{4} \right)^{1/3},
\label{Trelation}
\end{align}
where $g_{\ast}(t)$ is the effective number of degrees of freedom in the plasma.

Next, we can extrapolate the properties of neutrinos in the present universe. Since the present CMB temperature is observed to be $T_{\gamma}(t_0)\simeq2.7255\ {\rm K}$ \cite{Fixsen:2009ug},
the present neutrino temperature $T_{\nu}(t_0)$ and the present neutrino distribution function $f_{0}(\bm{p})$ are evaluated through Eq.~(\ref{Trelation}) as
\begin{align}
T_{\nu}(t_0)&\simeq1.9454\ {\rm K}, \label{PNT} \\
f_{0}(\bm{p})&=\frac{1}{e^{|\bm{p}|/T_{\nu}(t_0)}+1}.
\label{DFI}
\end{align}
Using Eqs.~(\ref{ND}) and (\ref{PNT}), the current neutrino number density per one degree of freedom is estimated as
\begin{align}
n_0=\frac{3\zeta(3)}{4\pi^2}T_{\nu}(t_0)^3 \simeq 56.01\ {\rm cm^{-3}}.
\end{align}
We can also calculate the average magnitude of a cosmic neutrino's momentum in the present universe,
\begin{equation}
\langle p_0 \rangle =\frac{1}{n_0}\int\frac{d^3p}{(2\pi)^3}|\bm{p}|f_0(\bm{p}) \approx 3.15 T_\nu(t_0) \approx 5.3\times 10^{-4}\mathrm{eV}.
\end{equation}
Since $\langle p_0 \rangle$ is much smaller than $\sqrt{\Delta m_{21}^2}$ and $\sqrt{|\Delta m_{3l}^2|}\ (l=1,2)$, where $\Delta m_{ij}^2=m_i^2-m_j^2$ are the mass squared differences of two neutrino species, at least two mass-eigenstates of neutrinos are non-relativistic today.
From this, it is easier to follow the evolution of neutrinos in the mass-diagonal basis because we can easily quantize non-relativistic neutrinos and calculate this capture rate.

Finally, we consider the history of neutrino spin states and which neutrino spin states are populated in the present universe.
In the early universe, since the weak interaction is chiral, only left-handed neutrinos and right-handed anti-neutrinos were produced in thermal equilibrium with the other standard model particles.  
On the other hand, since right-handed neutrinos and left-handed anti-neutrinos cannot interact with other particles via the weak interaction, we call these neutrinos sterile. Since these sterile neutrinos could not be produced in thermal equilibrium through the weak interaction, we assume that their number densities are negligibly small.

In the early universe, the chirality states of these ultra-relativistic neutrinos are conserved owing to the negligibility of their masses. However, in the current universe, the chirality states of neutrinos are not necessarily conserved since some neutrinos are non-relativistic. In this epoch, it is easier to follow the evolution of helicity states of neutrinos since their helicity states are conserved while non-relativistic neutrinos are freely streaming. The helicity of a particle is defined by the projection of its spin vector onto the direction of its momentum.

Thus, in the present universe, left-helical neutrinos (right-helical neutrinos) would be populated, which coincide with left-handed neutrinos (right-handed anti-neutrinos for Dirac type and right-handed neutrinos for Majorana type) in the early universe.

So, if neutrinos are Dirac fermions and neglecting the possible mixing of neutrino helicity (discussed further in section 2.3), the number density for each spin state in the current universe is
\begin{align}
n_{\nu_l}&=n_{\bar{\nu}_r}= n_0, \nonumber \\
n_{\nu_r}&\approx n_{\bar{\nu}_l} \approx 0,
\label{DN}
\end{align}
where $\nu_l$ ($\bar{\nu}_r$) denotes left-helical neutrinos (right-helical anti-neutrinos) while $\nu_r$ ($\bar{\nu}_l$) denotes right-helical sterile neutrinos (left-helical sterile anti-neutrinos). The present distribution function for each spin state of Dirac neutrino is
\begin{align}
f_{\nu_l}&=f_{\bar{\nu}_r}=f_{0}(\bm{p}), \nonumber \\
f_{\nu_r}&\approx f_{\bar{\nu}_l} \approx 0.
\end{align}

If neutrinos are Majorana fermions, there is no distinction between neutrinos and anti-neutrinos, and the lepton number is violated. In addition, Majorana sterile neutrinos are typically much heavier than active neutrinos through the see-saw mechanism \cite{Minkowski:1977sc, Yanagida:1979as, GellMann:1980vs, Mohapatra:1979ia, Schechter:1980gr} and completely decay into other particles in the early universe. Then the number density for each spin state in the current universe is
\begin{align}
n_{\nu_l}&=n_{\nu_r}=n_0, \nonumber \\
n_{N_r}&= n_{N_l}= 0,
\label{MN}
\end{align}
where $\nu_l$ ($\nu_r$) denotes left-helical neutrinos (right-helical neutrinos) while $N_r$ ($N_l$) denotes right-helical sterile neutrinos (left-helical sterile neutrinos).
In terms of the present distribution function, each spin state of Majorana neutrino has
\begin{align}
f_{\nu_l}&=f_{\nu_r}=f_{0}(\bm{p}), \nonumber \\
f_{N_r}&= f_{N_l}= 0.
\end{align}


\subsection{Neutrino spectral distortion in the neutrino decoupling}

In the previous section, we assumed that all neutrinos instantaneously stopped interacting with other particles. However, since the decoupling time actually depends on the momenta of neutrinos, neutrinos decoupled gradually. In particular, the decoupling temperature of neutrinos and the temperature of annihilation of $e^{\pm}$-pairs are so close that some $e^{\pm}$-pairs annihilate into neutrinos, injecting their energies into neutrinos. These annihilation processes become more efficient for neutrinos with higher energies because the interaction rates of relativistic particles with higher energies are larger \cite{Dicus:1982bz, Dolgov:2002wy}. Due to this energy injection, the neutrinos' distribution function after decoupling is distorted as
\begin{align}
f_{\nu_i}^d(\bm{p},t)=\frac{1}{e^{|\bm{p}|/\bar{T}_{\nu}(t)}+1}\bigl(1+\delta \bar{f}^d_{\nu_i}(\bm{p}, t) \bigl),
\end{align}
where $i$ denotes the mass-eigenstate of a neutrino.
Since the energy injection into neutrinos makes the decrease of the neutrino temperature effectively slower, the ratio of the photon temperature, $T_{\gamma}$, to the actual neutrino temperature, $\bar{T}_{\nu}$, is smaller, and it is given by the latest calculation in \cite{Akita:2020szl},\footnote{Since the actual neutrino spectrum also deviates from the Fermi-Dirac distribution, the neutrino temperature is not uniquely defined. Here we determine the neutrino temperature by requiring that it simply decreases in proportion to the inverse of the scale factor while the photon temperature does not due to the $e^{\pm}$-pair annihilation and the non-instantaneous decoupling effects.}
\begin{align}
\frac{T_{\gamma}(t_0)}{\bar{T}_{\nu}(t_0)}&=1.39797, \nonumber \\
\bar{T}_{\nu}(t_0)&=1.9496\ {\rm K}.
\end{align}
Hereafter we only consider the quantities in the present universe. These corrections to the Fermi-Dirac distribution for the mass-eigenstates of neutrinos in the SM are studied in Fig. 5 of ref. \cite{Akita:2020szl}. These corrections also modify the number densities of neutrinos in the current universe, which affect the neutrino capture rate on tritium. Including these corrections, the number densities of neutrinos in the current universe are given by
\begin{align}
n_{\nu_i}^d=\bar{n}_0\left(1+\delta \bar{n}_{\nu_i}^d \right),
\end{align}
where
\begin{align}
\bar{n}_0=\frac{3\zeta(3)}{4\pi^2}\bar{T}_{\nu}(t_0)^3 \simeq 56.376\ {\rm cm^{-3}}.
\end{align}
This difference in neutrino temperature, when compared to the instantaneous decoupling limit, induces a change of $0.65\%$ in the current neutrino number density. The values of $\delta \bar{n}_i$ in the SM are listed in Table~\ref{tb:ND}. These values do not depend on the neutrino mass ordering.  We can also parametrize the deviation of the current distribution functions and number densities from those in the instantaneous decoupling limit as
\begin{align}
f_{\nu_i}^d(\bm{p},t_0)&=f_0(\bm{p})\bigl(1+\delta f^d_{\nu_i}(\bm{p}, t_0) \bigl), \nonumber \\
n_{\nu_i}^d&=n_0\left(1+\delta n_{\nu_i}^d \right).
\end{align}
The values of $n_{\nu_i}^d$ and $\delta n_{\nu_i}^d$ are listed in Table~\ref{tb:ND2}.

Detecting these corrections to neutrino number densities would reveal not only precise number densities but also the contribution of neutrinos to $N_{\rm eff}$. Thus, the detection of the distortions of neutrino spectra and number densities will enable us to distinguish between models of the early universe more precisely.

\begin{table}[h]
\begin{center}
	\begin{tabular}{|c|c|c|}
		\hline
		 $\delta \bar{n}_{\nu_1}^d$ (\%) & $\delta \bar{n}_{\nu_2}^d$ (\%) & $\delta \bar{n}_{\nu_3}^d$ (\%)   \\
		\hline
		 0.468 & 0.350 & 0.248  \\
		\hline
	\end{tabular}
	\caption{The deviation of present number densities of neutrinos from the Fermi-Dirac distribution $[e^{|\bm{p}|/\bar{T}(t_0)}+1]^{-1}$ in the mass basis in the SM \cite{Akita:2020szl}.}
  \label{tb:ND}
\end{center}
\end{table}

\begin{table}[h]
\begin{center}
	\begin{tabular}{|c|c|c|c|c|c|c|}
		\hline
	  $\delta n_{\nu_1}^d$ (\%) & $\delta n_{\nu_2}^d$ (\%) & $\delta n_{\nu_3}^d$ (\%) & $n_{\nu_1}^d\ (\mathrm{cm}^{-3})$  & $ n_{\nu_2}^d\ (\mathrm{cm}^{-3})$ & $n_{\nu_3}^d\ (\mathrm{cm}^{-3})$  \\
		\hline
		 1.13 & 1.01 & 0.91 & 56.64 & 56.57 & 56.52  \\
		\hline
	\end{tabular}
	\caption{The present number densities of neutrinos and the deviation of those from the instantaneous decoupling limit in the mass basis including neutrino spectral distortions in the SM.}
  \label{tb:ND2}
\end{center}
\end{table}


\subsection{Gravitational clustering}

After the decoupling, neutrinos freely streamed until today. However, near the Earth, non-relativistic neutrinos can cluster locally in the gravitational potential of our Galaxy and nearby galaxies. Due to this clustering, the local distribution function is modified and the local number density is enhanced when compared with the global distribution function and number density, which also enhance the capture rate of cosmic neutrinos. The local distribution function and number density in the present universe are parametrized as
\begin{align}
f_{\nu_i}(\bm{p},t_0)&=f_{\nu_i}^d(\bm{p},t_0)\bigl(1+\delta f_{\nu_i}^c(\bm{p}, t_0) \bigl), \nonumber \\
n_{\nu_i}&=n_{\nu_i}^d\left(1+\delta n_{\nu_i}^c \right).
\label{CFN}
\end{align}
Using the linear approximation, Eq.~(\ref{CFN}) in the present universe can be rewritten as
\begin{align}
f_{\nu_i}(\bm{p},t_0)&\simeq f_0(\bm{p})\bigl(1+\delta f_{\nu_i}^c(\bm{p}, t_0) + \delta f_{\nu_i}^d(\bm{p}, t_0)\bigl), \nonumber \\
n_{\nu_i}&\simeq n_0\left(1+\delta n_{\nu_i}^c+\delta n_{\nu_i}^d\right).
\end{align}
Note that $\delta f_{\nu_i}^c$ has not yet been estimated in previous works, although we can in principle estimate this modification in the same way as one estimates the enhancement of the neutrino number density. We too leave this estimation of $\delta f_{\nu_i}^c$ to future work. $\delta n^c_{\nu_i}$ was estimated by a method to solve the collisionless Boltzmann equation for a system including cold dark matter halos and neutrinos in ref.~\cite{Singh:2002de}, and by a method called N-one-body simulations in refs.~\cite{Ringwald:2004np, deSalas:2017wtt, Zhang:2017ljh, Mertsch:2019qjv}. Through the latter method, they computed the time evolution of trajectories of $N$-independent neutrinos in the gravitational potential of the Milky Way, Virgo cluster, and Andromeda galaxy in the latest calculation \cite{Mertsch:2019qjv}. This study shows that the difference between a Navarro-Frenk-White(NFW) and an Einasto profile for the dark matter around our galaxy is negligibly small. For the case of the SM neutrinos, $\delta n_{\nu_i}^c$ have been also calculated in \cite{Mertsch:2019qjv} and we display some of these values in Table~\ref{tb:CND} for reference. From Tables~\ref{tb:ND2} and~\ref{tb:CND}, one can observe that the clustering effect of the mass-eigenstates of neutrinos with $m\simeq 50\ {\rm meV}$ is dominant compared to the non-thermal effect in the early universe, while the clustering effect of neutrinos with $m\simeq 10\ {\rm meV}$ is slightly less than the non-thermal effect in the early universe.
However, since the lightest mass of neutrino has not yet been determined, the lightest neutrino could be light enough not to cluster well, and the non-thermal effect in the early universe can be dominant compared to the clustering effect for the lightest neutrinos. Thus, it may be easier to detect the non-thermal neutrino spectral distortion through the capture of the lightest neutrino species using the tritium target, albeit requiring an extremely good energy resolution (approximately $0.4$ meV, shown in section 4.2).

In addition, gravitational clustering of massive neutrinos may induce the mixing of neutrino helicity \cite{Long:2014zva, Roulet:2018fyh, Duda:2001hd}, although the quantitative calculation has also not yet been achieved. Since neutrinos orbit around our Galaxy in the gravitational potential, its direction of momentum would change whereas its spin does not, which would induce the change of its helicity. If the helicities of neutrinos change completely, the distribution functions and number densities of all Dirac neutrinos become $f_{\nu_l}= f_{\bar{\nu}_r}= f_{\nu_r} = f_{\bar{\nu}_l}=  f_{\nu_i}/2$ and $n_{\nu_l}=n_{\bar{\nu}_r}= n_{\nu_r} = n_{\bar{\nu}_l}=n_{\nu_i}/2$, respectively. On the other hand, since the helicities of Majorana neutrinos initially mixed completely, the distribution functions and number densities of Majorana neutrinos remain unchanged. In spite of the possible helicity flipping for massive neutrinos, the capture rate would not change much since this capture rate depends mainly on the number density summed over helicities at leading order. Therefore, in the following, we will neglect the effect of helicity flipping for the estimation of the capture rate.

\begin{table}[h]
\begin{center}
	\begin{tabular}{|c|c|c|c|c|c|c|}
		\hline
		$m$ ({\rm meV}) & $\delta n^c$ (\%) & $n\ (\mathrm{cm}^{-3})$   \\
		\hline
		10 & 0.53 & 56.31  \\
		50 & 12 &  62.73 \\
		\hline
	\end{tabular}
	\caption{The current number densities of neutrinos and the deviation of those from the instantaneous decoupling limit with a mass $m$ including clustering effect by our Galaxy and not including neutrinos' spectral distortions \cite{Mertsch:2019qjv}.}
  \label{tb:CND}
\end{center}
\end{table}


\section{Precise neutrino capture rate including cosmological effects}
\label{sec:3}

In this section, including cosmological effects such as neutrino spectral distortion from the neutrino decoupling and gravitational clustering, we formulate the expected capture rate of neutrinos from the C$\nu$B on a tritium target through the following process,
\begin{equation}
\nu_i + \mathrm{^3H} \rightarrow \mathrm{^3He}+e^-,
\label{IBD}
\end{equation}
where $\nu_i$ denotes a neutrino in the mass-diagonal basis. In this calculation, we take the neutrino velocity $v_{\nu}$  into account, which contributes to the capture rate as the next-to-leading order effect and is comparable with these cosmological effects. In addition, the full expression of the capture rate at the tree-level is shown in appendix~\ref{appa}.

In order to derive the neutrino capture rate including such cosmological effects, we begin by considering the scattering amplitude for the process in Eq. (\ref{IBD}). 
Since we are interested in the reaction at an energy much lower than the weak boson masses, the approximate four-Fermi interaction process can be used to calculate the amplitude. In this case, the matrix element is given by
\begin{align}
i\mathcal{M}_i =-i \frac{G_F}{\sqrt{2}}V_{ud}U_{ei}^{\ast}\biggl[\bar{u}_e\gamma^\mu(1-\gamma^5)u_{\nu_i}\biggl]\biggl[\bar{u}_{\mathrm{^3H}}\gamma_\mu\left(\langle f_F \rangle - \frac{g_A}{\sqrt{3}g_V}\langle g_{GT}\rangle\gamma^5\right)u_{\mathrm{^3He}}\biggl],
\end{align}
where $u_{\alpha}$ denotes the Dirac spinor for species $\alpha$, $V_{ud}\simeq 0.9740$ \cite{Zyla:2020zbs} is a component of the Cabibbo-Kobayashi-Maskawa (CKM) matrix, and $U_{ei}$ is an element of the Pontecorvo-Maki-Nakagawa-Sakata (PMNS) matrix. $g_A\simeq1.2723$ and $g_V\simeq1$ are the axial and vector coupling constants respectively, and $\langle f_F \rangle\simeq 0.9998$ and $\langle g_{GT} \rangle \simeq \sqrt{3} \times (0.9511 \pm 0.0013)$ denote the reduced matrix elements of the Fermi and Gamow-Teller (GT) operators respectively \cite{Baroni:2016xll}.
The above value of $\langle g_{GT} \rangle$ is estimated through the observation of the tritium half-life and the value of $\langle f_F \rangle$.
Although the uncertainty of this ``experimental'' value is $0.1\%$, the theoretical calculation of $\langle g_{\rm GT} \rangle$ still includes an uncertainty of a few $\%$ \cite{Baroni:2016xll}.

In the inverse $\beta$-decay experiment, the spins of the outgoing electron and nucleus would not be measured. In addition, the spin of the initial nucleus would not be identified either. However, particularly in the case of Dirac neutrino, the initial number density for each spin state of neutrino can be seen in Eq.~(\ref{DN}), although the number density for each spin state of Majorana neutrino is the same. For these reasons, we calculate the squared scattering amplitude summed over the spin of $e$ and $\mathrm{^3He}$ and averaged over $\mathrm{^3H}$. In the rest frame of $\mathrm{^3H}$, the result is
\begin{align}
&\frac{1}{2}\sum_{s_e,s_{\mathrm{^3He}},s_{\mathrm{^3H}}=\pm\frac{1}{2}}|\mathcal{M}|_i^2(s_{\nu}) \nonumber \\
&=8G_F^2|V_{ud}|^2|U_{ei}|^2m_\mathrm{^3He}m_\mathrm{^3H}E_eE_{\nu_i} \nonumber \\
&\ \ \ \ \times \biggl[(1-2s_\nu v_{\nu_i})\left(\langle f_F \rangle^2+\frac{g_A^2}{g_V^2}\langle g_{GT} \rangle^2 \right)
+(v_{\nu_i}-2s_\nu)\cos \theta v_e \left( \langle f_F \rangle^2-\frac{g_A^2}{3g_V^2}\langle g_{GT} \rangle^2 \right)\biggl],
\label{M^2}
\end{align}
where $m_{\mathrm{^3He}} \simeq 2808.391\ {\rm MeV}$ and $m_{\mathrm{^3H}} \simeq 2808.921\ {\rm MeV}$ are the nuclear masses\footnote{The nuclear masses $m_{\mathrm{^3He}}$ and $m_{\mathrm{^3H}}$ are obtained from the atomic masses $M_{\mathrm{^3He}}\simeq 2809.413\ {\rm MeV}$ and $M_{\mathrm{^3H}}\simeq 2809.432\ {\rm MeV}$ \cite{Wang2016}, using the following relations, $m_{\mathrm{^3He}}=M_{\mathrm{^3He}}-2m_e+24.58678\ {\rm eV}$ and $m_{\mathrm{^3H}}=M_{\mathrm{^3H}}-m_e+13.59811\ {\rm eV}$.
The last values on the right hand side in the previous two equations represent the atomic binding energies.}
of the $\mathrm{^3He}$ and $\mathrm{^3H}$ respectively,
$s_\alpha=\frac{1}{2}\ (-\frac{1}{2})$ denotes the right (left) helicity for species $\alpha$ , $v_\alpha=\frac{|\bm{p}_\alpha|}{E_\alpha}$ is the velocity for species $\alpha$, and $\cos\theta=\frac{\bm{p}_e\cdot\bm{p}_{\nu}}{|\bm{p}_e||\bm{p}_{\nu}|}$ is the angle between the electron and neutrino momenta. 

We derive the differential cross section from Eq.~(\ref{M^2}) (see also appendix.~\ref{appa}) up to the next-to-leading order as
\begin{align}
\frac{d\sigma_i(s_{\nu})}{d\cos\theta} &=\frac{G_F^2}{4\pi}|V_{ud}|^2|U_{ei}|^2\frac{m_\mathrm{^3He}}{m_\mathrm{^3H}v_{\nu_i}}F(2,E_e)E_e|\bm{p}_e| \nonumber \\
&\ \ \ \ \times \biggl[(1-2s_\nu v_{\nu_i})\left(\langle f_F \rangle^2+\frac{g_A^2}{g_V^2}\langle g_{GT} \rangle^2 \right) +(v_{\nu_i}-2s_\nu)\cos \theta v_e \left( \langle f_F \rangle^2-\frac{g_A^2}{3g_V^2}\langle g_{GT} \rangle^2 \right)\biggl],
\end{align}
where $F(Z,E_e)$ is the Fermi function expressed as
\begin{align}
F(Z,E_e)=\frac{2\pi\alpha ZE_e/|\bm{p}_e|}{1-e^{-2\pi\alpha ZE_e/|\bm{p}_e|}}.
\end{align}
This function represents an enhancement factor by an Coulombic attraction of the out going electron and proton \cite{Primakoff:1959chj}. $Z$ is the atomic number of the daughter nucleus and $Z=2$ in our case. $\alpha\simeq 1/137.036$ is the fine structure constant. After integrating over the angle $\theta$, the total cross section multiplied by the neutrino velocity is given by 
\begin{align}
\sigma_i(s_\nu)v_{\nu_i} 
&=\frac{G_F^2}{2\pi}|V_{ud}|^2|U_{ei}|^2\frac{m_\mathrm{^3He}}{m_\mathrm{^3H}}F(2,E_e)E_e|\bm{p}_e| \nonumber \\
&\ \ \ \ \times(1-2s_\nu v_{\nu_i})\left(\langle f_F \rangle^2+\frac{g_A^2}{g_V^2}\langle g_{GT} \rangle^2 \right).
\label{sigma}
\end{align}
Although $|s_{\nu}v_{\nu}|\ll1$ for non-relativistic neutrinos, the cross section for each helicity state of neutrino is slightly different. 
Then, the cross sections for Dirac and Majorana neutrinos also are different since the abundance of the helicity states of these neutrinos are different as in Eqs. (\ref{DN}) and (\ref{MN}).

We now calculate the total capture rate of cosmic neutrinos $\Gamma_{\rm{C\nu B}}$ for some tritium sample with $N_T=\frac{M_T}{M_{\mathrm{^3H}}}$ particles, where $M_T$ is the total mass of the experimental setup of tritium and $M_{\mathrm{^3H}} \simeq 2809.432\ {\rm MeV}$ is the atomic mass of tritium \cite{Wang2016}.
 This total capture rate $\Gamma_{\rm{C\nu B}}$ can be rewritten as
\begin{align}
\Gamma_{\rm{C\nu B}} = \sum_{i=1}^{N_{\nu}}\Gamma_i,
\end{align}
where $N_{\nu}$ is the number of (mass) species of neutrinos. $\Gamma_i$ is the total capture rate of a given mass-eigenstate of neutrino $\nu_i$, given by
\begin{align}
\Gamma_i = N_T \sum_{s_{\nu}=\pm\frac{1}{2}}\int\frac{d^3p_{\nu}}{(2\pi)^3} \sigma_i(\bm{p}_{\nu},s_\nu)v_{\nu_i}f_{\nu_i}(\bm{p}_{\nu},s_{\nu}),
\label{Gi}
\end{align}
where $f_{\nu_i}(p_{\nu},s_\nu)$ is the distribution function for $\nu_i$ in the present universe. Plugging Eq.~(\ref{sigma}) into Eq.~(\ref{Gi}) yields
\begin{align}
\Gamma_i&=N_T\frac{G_F^2}{2\pi}|V_{ud}|^2|U_{ei}|^2\frac{m_\mathrm{^3He}}{m_\mathrm{^3H}}\left(\langle f_F \rangle^2+\frac{g_A^2}{g_V^2}\langle g_{GT} \rangle^2 \right) \nonumber \\
&\ \ \ \ \times\sum_{s_{\nu}=\pm\frac{1}{2}}\int\frac{d^3p_{\nu}}{(2\pi)^3}\ f_{\nu_i}(\bm{p}_{\nu},s_\nu) F(2,E_e)E_e|\bm{p}_e|(1-2s_\nu v_{\nu_i}),
\label{Gammai}
\end{align}
where, for left-helical Dirac neutrinos, left-helical Majorana neutrinos and right-helical Majorana neutrinos 
\begin{align}
f_{\nu_i}(\bm{p}_{\nu},s_{\nu})=f_0(\bm{p}_{\nu})\bigl(1+\delta f_{\nu_i}^c(\bm{p}_{\nu}, t_0) + \delta f_{\nu_i}^d(\bm{p}_{\nu}, t_0)\bigl),
\end{align}
and for other sterile neutrinos
\begin{align}
f_{\nu_i}(\bm{p}_{\nu},s_{\nu})= 0.
\end{align}
Here $f_0(\bm{p})$ is the current neutrino distribution function in the instantaneous decoupling limit, defined as Eq.~(\ref{DFI}).
It should be again noted that we have neglected the possible helicity flip effects for massive neutrinos by the neutrino clustering since the helicity-dependent part of $\Gamma_i$ is already suppressed by $v_{\nu_i}$.

The energy and momentum of an electron in Eq.~(\ref{Gammai}) depend on the neutrino masses and energies because of energy-momentum conservation. In the rest frame of $\mathrm{^3H}$, the electron energy and momentum are written as (see appendix.~\ref{appb})
\begin{align}
E_e &\simeq  K_{\rm end}^0 + m_e + E_{\nu_i}, \nonumber \\
|\bm{p}_e| &= \sqrt{E_e^2-m_e^2},
\label{Ee}
\end{align}
where $K_{\rm end}^0$ is the beta decay endpoint kinetic energy for massless neutrinos given by
\begin{align}
K_{\rm end}^0= \frac{(m_{\mathrm{^3H}}-m_e)^2-m_{\mathrm{^3He}}^2}{2m_{\mathrm{^3H}}}.
\end{align}
The average value of $E_{\nu_i}$ is so small compared to $K_{\rm end}^0$ and $m_e$ that we can safely neglect this dependence in Eq.~(\ref{Gammai}).

In particular, the correction to the total capture rate from the cosmological effects, $\delta \Gamma_i$, is given by
\begin{align}
\delta \Gamma_i &=N_T\frac{G_F^2}{2\pi}|V_{ud}|^2|U_{ei}|^2\frac{m_\mathrm{^3He}}{m_\mathrm{^3H}} \left(\langle f_F \rangle^2+\frac{g_A^2}{g_V^2}\langle g_{GT} \rangle^2 \right) \nonumber \\
&\ \ \ \ \times \sum_{s_{\nu}=\pm\frac{1}{2}}\int\frac{d^3p_{\nu}}{(2\pi)^3}\ f_0(\bm{p}_\nu)(\delta f_{\nu_i}^c+\delta f^d_{\nu_i}) F(2,E_e)E_e|\bm{p}_e|(1-2s_\nu v_{\nu_i}).
\end{align}

In Eq.~(\ref{Ee}), the contribution of the neutrino momentum is very small, roughly $\langle p_0 \rangle/ m_e\times m_e\sim 10^{-9} m_e$. When we neglect the neutrino momentum in Eq.~(\ref{Ee}), Eq.~(\ref{Gammai}) reduces to a much simpler form with 
\begin{align}
\Gamma_i&\simeq N_T\frac{G_F^2}{2\pi}|V_{ud}|^2|U_{ei}|^2\frac{m_\mathrm{^3He}}{m_\mathrm{^3H}}\left(\langle f_F \rangle^2+\frac{g_A^2}{g_V^2}\langle g_{GT} \rangle^2 \right) \nonumber \\
&\ \ \ \ \times F(2,\tilde{E}_e)\tilde{E}_e|\tilde{\bm{p}}_e| \sum_{s_{\nu}=\pm\frac{1}{2}}\left(n_{\nu_i}-2s_{\nu}\langle v_{\nu_i} \rangle \right),
\label{SGammai}
\end{align}
where $\langle v_{\nu_i} \rangle$ is the (unnormalized) average magnitude of velocity for $\nu_i$ given by
\begin{align}
\langle v_{\nu_i} \rangle &= \int\frac{d^3p_{\nu}}{(2\pi)^3}\ f_{\nu_i}(\bm{p}_{\nu},s_\nu)v_{\nu_i},
\end{align}
and
\begin{align}
\tilde{E}_e&=K_{\rm end}^0+m_e + m_{\nu_i}, \nonumber \\
|\tilde{\bm{p}}_e|&=\sqrt{\tilde{E}_e^2 - m_e^2}.
\label{EcnuB}
\end{align}
The corrections to the total capture rate from cosmological effects also become
\begin{align}
\delta \Gamma_i&=N_T\frac{G_F^2}{2\pi}|V_{ud}|^2|U_{ei}|^2\frac{m_\mathrm{^3He}}{m_\mathrm{^3H}} \left(\langle f_F \rangle^2+\frac{g_A^2}{g_V^2}\langle g_{GT} \rangle^2 \right)  \nonumber \\
&\ \ \ \ \times F(2,\tilde{E}_e)\tilde{E}_e|\tilde{\bm{p}}_e|
\sum_{s_{\nu}=\pm\frac{1}{2}}( \delta n_{\nu_i}  -2s_\nu \langle \delta v_{\nu_i} \rangle),
\label{DeltaSGammai}
\end{align}
where
\begin{align}
 \delta n_{\nu_i} &=  \delta n^d_{\nu_i}  + \delta n_{\nu_i}^c , \nonumber \\
\langle \delta v_{\nu_i} \rangle &= \int\frac{d^3p_{\nu}}{(2\pi)^3}\ f_0(\bm{p}_\nu)\left(\delta f^d_{\nu_i}+\delta f_{\nu_i}^c \right)v_{\nu_i}.
\label{DeltaGammai}
\end{align}
We can represent Eq.~(\ref{DeltaGammai}) as a linear combination of the spectral distortion in the neutrino decoupling and the contribution from gravitational clustering,
\begin{align}
\delta \Gamma_i &= \delta \Gamma_i^d + \delta \Gamma_i^c, \nonumber \\
\delta \Gamma_i^d&=N_T\frac{G_F^2}{2\pi}|V_{ud}|^2|U_{ei}|^2\frac{m_\mathrm{^3He}}{m_\mathrm{^3H}}\left(\langle f_F \rangle^2+\frac{g_A^2}{g_V^2}\langle g_{GT} \rangle^2 \right) \nonumber \\
&\ \ \ \ \times F(2,\tilde{E}_e)\tilde{E}_e|\tilde{\bm{p}}_e| 
\sum_{s_{\nu}=\pm\frac{1}{2}}( \delta n_{\nu_i}^d  -2s_\nu \langle \delta v_{\nu_i}^d \rangle) , \nonumber \\
\delta \Gamma_i^c&=N_T\frac{G_F^2}{2\pi}|V_{ud}|^2|U_{ei}|^2\frac{m_\mathrm{^3He}}{m_\mathrm{^3H}} \left(\langle f_F \rangle^2+\frac{g_A^2}{g_V^2}\langle g_{GT} \rangle^2 \right)  \nonumber \\
&\ \ \ \ \times F(2,\tilde{E}_e)\tilde{E}_e|\tilde{\bm{p}}_e|
\sum_{s_{\nu}=\pm\frac{1}{2}}( \delta n_{\nu_i}^c  -2s_\nu \langle \delta v_{\nu_i}^c \rangle),
\end{align}
where
\begin{align}
\langle \delta v_{\nu_i}^d \rangle &= \int\frac{d^3p_{\nu}}{(2\pi)^3}\ f_0(\bm{p}_\nu)\delta f_{\nu_i}^d v_{\nu_i}, \nonumber \\
\langle \delta v_{\nu_i}^c \rangle &= \int\frac{d^3p_{\nu}}{(2\pi)^3}\ f_0(\bm{p}_\nu)\delta f_{\nu_i}^c v_{\nu_i}.
\end{align}

We comment on the order of each term in Eq.~(\ref{Gammai}) in terms of dimensionless parameters, $\delta f_{\nu_i}^c,\ \delta f_{\nu_i}^d,\ v_{\nu_i} $ and $|\bm{p}_e|^2/(m_{\mathrm{^3He}}E_e)$. 
Here, we assume $\delta f_{\nu_i}^d \sim \delta n_{\nu_i}^d$ and $\delta f_{\nu_i}^c \sim \delta n_{\nu_i}^c$ for the estimation of the order.
The value of $\delta f_{\nu_i}^c$ is $10^{-1}$ for $m_{\nu_i}\sim 50\ {\rm meV}$ and $5\times10^{-4}$ for $m_{\nu_i}\sim 10\ {\rm meV}$ as in Table~\ref{tb:CND}.
For $m_{\nu_i}< 10\ {\rm meV}$, we can neglect $\delta f_{\nu_i}^c$ when compared with the other dimensionless parameters. The value of $\delta f_{\nu_i}^d$ is $\mathcal{O}(10^{-2})$ as in Table~\ref{tb:ND2}.
If $m_{\nu_i}< 10\ {\rm meV}$, $\delta f_{\nu_i}^d$ is larger than $\delta f^c_{\nu_i}$. If $m_{\nu_i}\sim 10\ {\rm meV}$, $\delta f_{\nu_i}^c$ is comparable with $\delta f_{\nu_i}^d$.  Otherwise, $\delta f^c_{\nu_i}$ dominates over $\delta f_{\nu_i}^d$. The average value of $v_{\nu_i}\sim \langle p_0 \rangle/m_{\nu_i}$ is $10^{-2}$ for $m_{\nu_i}\sim 50\ {\rm meV}$ and $5\times 10^{-2}$ for $m_{\nu_i}\sim 10\ {\rm meV}$. In order to calculate the capture rate precisely, we should include the neutrino velocity $v_{\nu_i}$.

 In order to see easily and clearly the signature of the distortions owing to interactions in the early universe, we would need that their clustering effect is much smaller than the effect of the distortions, which is the case for $m_{\nu_i}< 10\ {\rm meV}$. Though we are unsure whether there exists a neutrino species with $m_{\nu_i}< 10\ {\rm meV}$, the lightest neutrinos can have such a tiny mass.
In particular, for the case of $m_{\nu_i}=0$ (or any extremely small mass), we can take $\delta f_{\nu_i}^c =0$ and $v_{\nu_i}=1$, and neglect the other smaller corrections. Then we get a simple form of the capture rate of the massless neutrino $\Gamma^{m_{\nu}=0}_i$:
\begin{align}
&\Gamma_i^{m_{\nu}=0} \nonumber \\
&\simeq N_T\frac{G_F^2}{\pi}|V_{ud}|^2|U_{ei}|^2\frac{m_\mathrm{^3He}}{m_\mathrm{^3H}}\left(\langle f_F \rangle^2+\frac{g_A^2}{g_V^2}\langle g_{GT} \rangle^2 \right)F(2,\tilde{E}_e^0)\tilde{E}_e^0|\tilde{\bm{p}}_e^0|(n_0+\delta n_i),
\end{align}
where
\begin{align}
\tilde{E}_e^0&= K_{\rm end}^0 + m_e, \nonumber \\
|\tilde{\bm{p}}_e^0| &= \sqrt{(\tilde{E}_e^0)^2-m_e^2}.
\end{align}
Note that in the case with $m_{\nu_i}=0$, only left-helical, that is, left-chiral (massless) neutrinos can be captured by tritium.


\section{Estimating the neutrino capture rate and the spectrum}
\label{sec:4}

\subsection{The capture rate}

In this section, we estimate the value of the neutrino capture rate on a $100$ gram tritium target, including cosmological effects.
Here we mainly focus on the case that the lightest neutrinos are (effectively) massless and hence there is no effect of gravitational clustering for them.
Under this assumption, we consider both the normal and inverted hierarchies of neutrino masses. In addition, we consider both Dirac and Majorana neutrinos. For massive neutrinos, we consider the capture rate including both the neutrino spectral distortion in the early universe and the gravitational clustering effect. For (effectively) massless neutrinos, we calculate the capture rate with the spectral distortion and without the clustering effect.
In the cases for massive neutrinos, we take $\langle v_{\nu_i} \rangle \simeq \langle v_{\nu_i}^0 \rangle =\int\frac{d^3p}{(2\pi)^3}\ f_0(\bm{p})v_{\nu_i}$ and $\langle \delta v_{\nu_i} \rangle \simeq 0$, which is the (unnormalized) average magnitude of velocity without the clustering effect and spectral distortion in the early universe because such effects on the average velocity correspond at most to the next-to-next-to leading order (NNLO). As repeated before, we also neglect the helicity flip effect by the neutrino clustering, which is also at most an NNLO effect. 

\subsubsection{The normal hierarchy case} 

First, the observed values of neutrino squared-mass differences from neutrino oscillation experiments are \cite{Esteban:2020cvm}
\begin{align}
\Delta m_{21}^2 \simeq (8.6\ {\rm meV})^2\ \ \ \ {\rm and}\ \ \ \ |\Delta m_{3l}^2|  \simeq (50\ {\rm meV})^2.
\end{align}
Due to the unknown sign of $\Delta m_{3l}^2$, two possible mass hierarchies are allowed. One of them is called the normal hierarchy:
\begin{align}
{\rm Normal\ hierarchy\ (NH)}:\ \Delta m_{31}^2>0,\ \ m_1<m_2<m_3,
\end{align}
where we define $\Delta m_{3l}^2=\Delta m_{31}^2$ as in ref. \cite{Esteban:2020cvm}.
In the normal hierarchy, $\nu_1$ can be massless. If we set $m_1=0$, we get the three masses of neutrinos as
\begin{align}
m_1\simeq0\ {\rm meV},\ \ m_2\simeq 8.6\ {\rm meV},\ \ {\rm and}\ \  m_3 \simeq 50\ {\rm meV}.
\end{align}
In this case, we calculate the capture rate for $\nu_2$ and $\nu_3$, including both the neutrino spectral distortion in the early universe and the gravitational clustering effect whereas the calculation of the capture rate for $\nu_1$ involves only the neutrino spectral distortion.

In the case of Majorana neutrinos, the total capture rate, $\Gamma_i^M$, its deviation originating from the spectral distortion from the decoupling, $\delta \Gamma_i^{Md}$, and that from the gravitational clustering effects, $\delta \Gamma_i^{Mc}$, are given by,
considering $100$ g of tritium,
\begin{align}
\Gamma_1^M\simeq 5.48\ {\rm yr^{-1}},  \ \ \ \  \Gamma_2^M\simeq 2.40\ {\rm yr^{-1}},  \ \ \ \   \Gamma_3^M\simeq 0.200 \ {\rm yr^{-1}},
\label{ValueM}
\end{align}
\begin{align}
\delta \Gamma_1^{Md}\simeq 0.061\ {\rm yr^{-1}},  \ \ \ \  \delta \Gamma_2^{Md}\simeq 0.024\ {\rm yr^{-1}},  \ \ \ \  \delta \Gamma_3^{Md}\simeq 1.6 \times 10^{-3} \ {\rm yr^{-1}},
\label{DGMd}
\end{align}
\begin{align}
\delta \Gamma_1^{Mc}\simeq 0\ {\rm yr^{-1}},  \ \ \ \  \delta \Gamma_2^{Mc}\simeq 0.013\ {\rm yr^{-1}},  \ \ \ \  \delta \Gamma_3^{Mc}\simeq 0.021 \ {\rm yr^{-1}},
\label{DGMc}
\end{align}
where we take the following values of the PMNS matrix,
\begin{align}
|U_{e1}|^2\simeq 0.681 ,\ \ \ \ |U_{e2}|^2\simeq 0.297,\ \ \ \ |U_{e3}|^2\simeq 0.0222.
\label{PMNS}
\end{align}
Note that the current errors of the PMNS matrix and the neutrino masses are about $10\%$ at 3$\sigma$ level \cite{Esteban:2020cvm}.
In particular, in the Majorana case, Eq.~(\ref{ValueM}) is insensitive to the neutrino mass except for the gravitational clustering effect because the mass-dependent terms, which corresponds the velocity-dependent terms in Eq.~(\ref{SGammai}), are canceled for the Majorana neutrinos by summing over helicities.
In order to observe the effects of Eqs.~(\ref{DGMd}) and (\ref{DGMc}), we need to have about $10^4$ events of cosmic neutrino capture since these cosmological effects modify the capture rates at a $1\%$ level.
For 100 grams of tritium, we cannot experimentally observe these effects since the half-life of tritium is 12.32 years.
To observe these contributions, we would need an experiment with about 10 kilograms of tritium.

In the case of Dirac neutrinos, the total capture rate, $\Gamma_i^D$, its deviation originating from the spectral distortion from the decoupling, $\delta \Gamma_i^{Dd}$, and that from the gravitational clustering effects, $\delta \Gamma_i^{Dc}$, are also given by,
considering $100$ g of tritium,
\begin{align}
\Gamma_1^D\simeq 5.48\ {\rm yr^{-1}},  \ \ \ \ \Gamma_2^D\simeq 1.27\ {\rm yr^{-1}},  \ \ \ \  \Gamma_3^D\simeq 0.101 \ {\rm yr^{-1}}, 
\end{align}
\begin{align}
\delta \Gamma_1^{Dd}\simeq 0.061\ {\rm yr^{-1}},  \ \ \ \  \delta \Gamma_2^{Dd}\simeq 0.012\ {\rm yr^{-1}},  \ \ \ \   \delta \Gamma_3^{Dd}\simeq 8.0 \times 10^{-4} \ {\rm yr^{-1}},
\end{align}
\begin{align}
\delta \Gamma_1^{Dc}\simeq 0\ {\rm yr^{-1}},  \ \ \ \  \delta \Gamma_2^{Dc}\simeq 6.3 \times 10^{-3}\ {\rm yr^{-1}},  \ \ \ \   \delta \Gamma_3^{Dc}\simeq 0.011\ {\rm yr^{-1}}.
\end{align}
Finally, the ratios between the capture rates for Dirac and Majorana neutrinos are
\begin{align}
\Gamma_1^M/\Gamma_1^D=1, \ \ \ \ \Gamma_2^M/\Gamma_2^D\simeq 1.89, \ \ \ \ \Gamma_3^M/\Gamma_3^D\simeq 1.98.
\end{align}

\subsubsection{The inverted hierarchy case}
The other possibility of neutrino mass hierarchy is the inverted ordering, which is
\begin{align}
{\rm Inverted\ hierarchy\ (IH)}:\ \Delta m_{32}^2<0,\ \ m_3<m_1<m_2,
\end{align}
where we define $\Delta m_{3l}^2=\Delta m_{32}^2$ \cite{Esteban:2020cvm}.
In the inverted hierarchy, $\nu_3$ can be massless. If we set $m_3=0$, we get the three masses of neutrinos as
\begin{align}
m_1\simeq49.3\ {\rm meV},\ \ m_2\simeq 50\ {\rm meV}\ \ {\rm and}\ \  m_3 \simeq 0\ {\rm meV}.
\end{align}

In the case of Majorana neutrinos, the total capture rate, $\Gamma_i^M$, its deviation originating from the spectral distortion from the decoupling, $\delta \Gamma_i^{Md}$, and that from the gravitational clustering effects, $\delta \Gamma_i^{Mc}$, are given by,
considering $100$ g of tritium,
\begin{align}
\Gamma_1^M\simeq 6.13\ {\rm yr^{-1}},  \ \ \ \  \Gamma_2^M\simeq 2.67\ {\rm yr^{-1}},  \ \ \ \   \Gamma_3^M\simeq 0.178 \ {\rm yr^{-1}},
\end{align}
\begin{align}
\delta \Gamma_1^{Md}\simeq 0.061\ {\rm yr^{-1}},  \ \ \ \  \delta \Gamma_2^{Md}\simeq 0.024\ {\rm yr^{-1}},  \ \ \ \  \delta \Gamma_3^{Md}\simeq 1.6 \times 10^{-3} \ {\rm yr^{-1}},
\end{align}
\begin{align}
\delta \Gamma_1^{Mc}\simeq 0.65\ {\rm yr^{-1}},  \ \ \ \  \delta \Gamma_2^{Mc}\simeq 0.28\ {\rm yr^{-1}},  \ \ \ \  \delta \Gamma_3^{Mc}\simeq 0 \ {\rm yr^{-1}},
\end{align}
where we also take the same values of the PMNS matrix as in Eq.~(\ref{PMNS}). Note that we take the same values of $\delta n_1^c$ for $m_1=49.3\ {\rm meV}$ and $\delta n_2^c$ for $m_2=50\ {\rm meV}$.

In the case of Dirac neutrinos, the total capture rate, $\Gamma_i^D$, its deviation originating from the spectral distortion from the decoupling, $\delta \Gamma_i^{Dd}$, and that from the gravitational clustering effects, $\delta \Gamma_i^{Dc}$, are also given by,
considering $100$ g of tritium,
\begin{align}
\Gamma_1^D\simeq 3.10\ {\rm yr^{-1}},  \ \ \ \ \Gamma_2^D\simeq 1.35\ {\rm yr^{-1}},  \ \ \ \  \Gamma_3^D\simeq 0.178 \ {\rm yr^{-1}},
\end{align}
\begin{align}
\delta \Gamma_1^{Dd}\simeq 0.031\ {\rm yr^{-1}},  \ \ \ \  \delta \Gamma_2^{Dd}\simeq 0.012\ {\rm yr^{-1}},  \ \ \ \   \delta \Gamma_3^{Dd}\simeq 1.6 \times 10^{-3}\ {\rm yr^{-1}},
\end{align}
\begin{align}
\delta \Gamma_1^{Dc}\simeq 0.33\ {\rm yr^{-1}},  \ \ \ \  \delta \Gamma_2^{Dc}\simeq 0.14\ {\rm yr^{-1}},  \ \ \ \   \delta \Gamma_3^{Dc}\simeq 0\ {\rm yr^{-1}}.
\end{align}
 The ratios between the capture rates for Dirac and Majorana neutrinos are
\begin{align}
\Gamma_1^M/\Gamma_1^D \simeq 1.98, \ \ \ \ \Gamma_2^M/\Gamma_2^D\simeq 1.98, \ \ \ \ \Gamma_3^M/\Gamma_3^D = 1.
\end{align}
Since the masses of $\nu_1$ and $\nu_2$ are almost the same, it is difficult to distinguish between these two signals. We discuss the possibility to distinguish between degenerate signals of two neutrino species in the next section.

\begin{figure}[htbp]
 \begin{minipage}{0.5\hsize}
  \begin{center}
   \includegraphics[width=85mm]{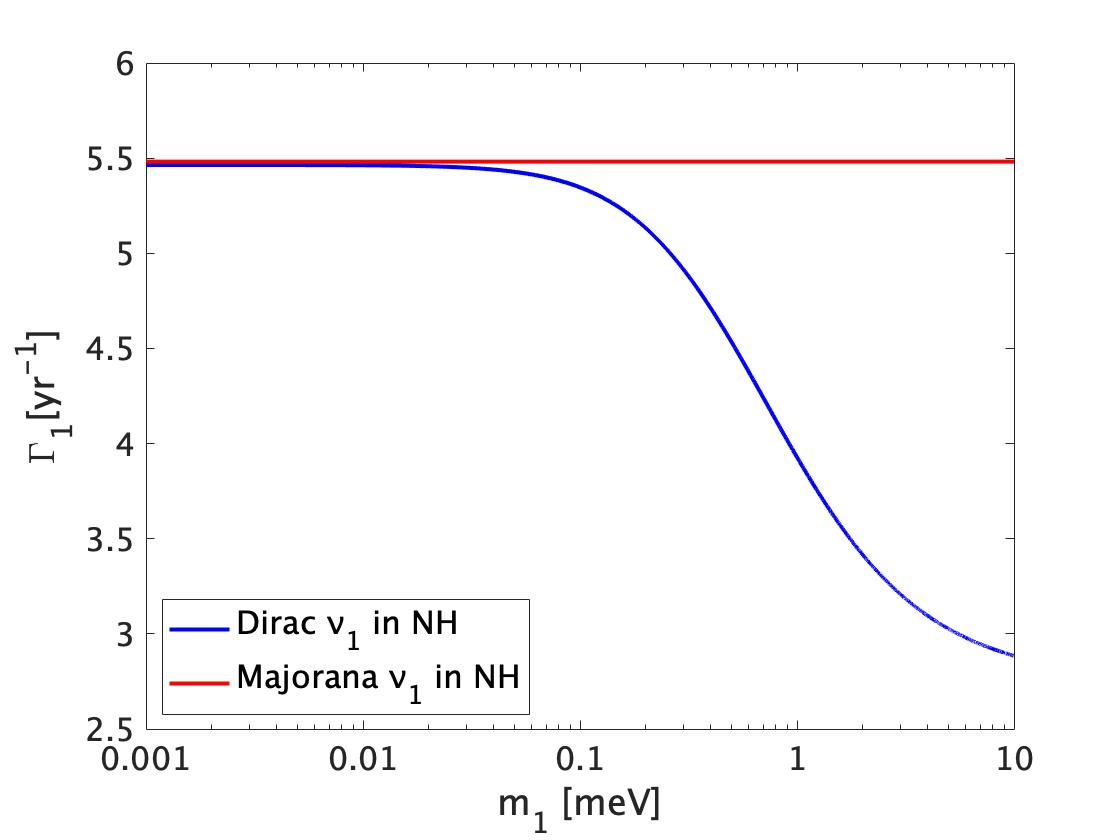}
  \end{center}
 \end{minipage}
 \begin{minipage}{0.5\hsize}
  \begin{center}
   \includegraphics[width=85mm]{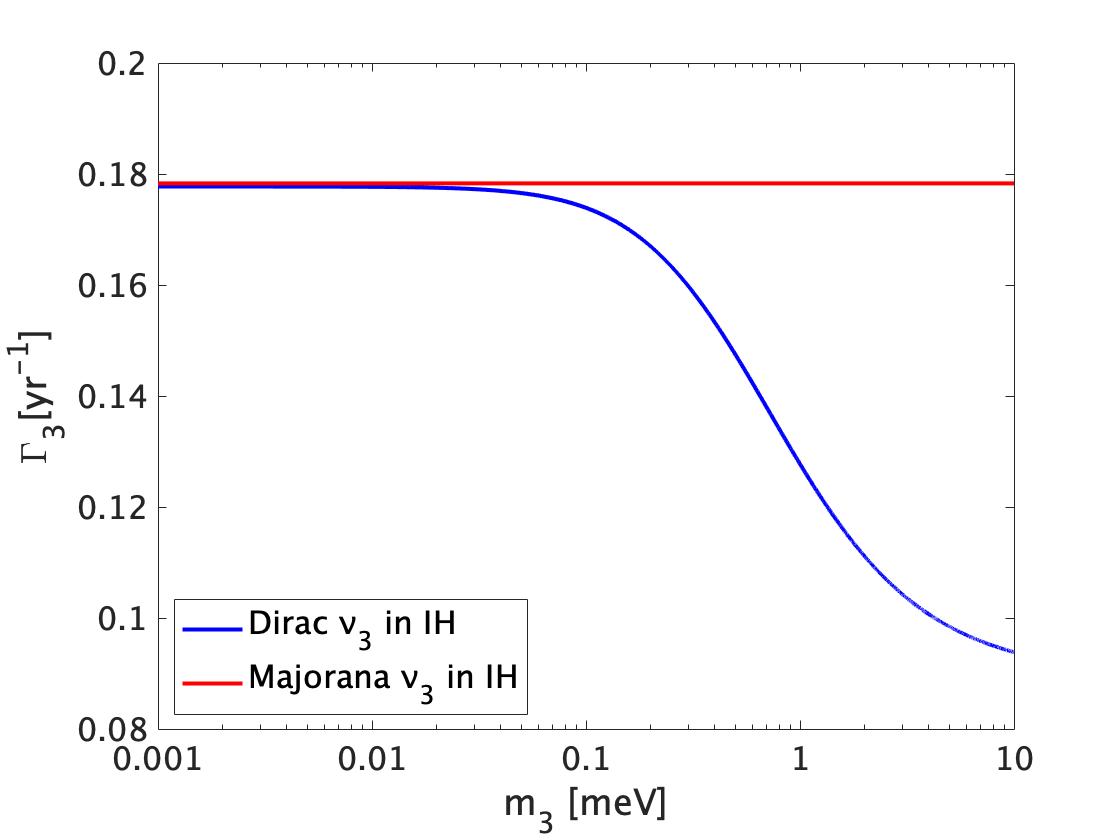}
  \end{center}
 \end{minipage}
  \vspace{-4mm}
 \caption{The capture rate for the lightest neutrinos on a 100 g tritium experiment in the NH (left panel) and IH (right panel) cases as a function of the lightest mass. We include the neutrino spectral distortion from the neutrino decoupling, using the results of ref.~\cite{Akita:2020szl}, but neglect the gravitational clustering effect since the lightest neutrinos are expected not to be affected by the gravitational clustering.}
 \label{fig:Gammanu}
\end{figure}

As pointed out in ref.~\cite{Roulet:2018fyh}, the capture rate for the lightest neutrino in the case of Dirac neutrinos significantly depends on the lightest mass through the (unnormalized) average magnitude of velocity, $\langle v_{\nu_i} \rangle$ in Eq.~(\ref{SGammai}) although, in the case of Majorana neutrinos, the dependence of velocity in Eq.~(\ref{SGammai}) is canceled due to the same populations of the left and right helical neutrinos. In Fig.~\ref{fig:Gammanu}, we show the capture rate for the lightest neutrinos in both hierarchies as a function of the lightest mass. From this figure, the capture rates for the lightest Dirac neutrinos with masses less than $0.01\ {\rm meV}$ are the same as those for the lightest Majorana neutrinos. On the other hand, the capture rates for Dirac neutrinos with masses more than $10\ {\rm meV}$ are almost half of those for Majorana neutrinos.

\subsection{The would-be spectra of an electron and the reconstruction of the spectrum of a cosmic neutrino}

Here, we first discuss the would-be observed spectrum of an electron emitted from the inverse $\beta$-decay process of tritium for the C$\nu$B. In particular, one of the main challenges for observing the signal of the C$\nu$B is the distinction of the signal from background. The main source of this background is tritium $\beta$-decay. Since tritium $\beta$-decay is a three-body process, $\mathrm{^3H}\rightarrow \mathrm{^3He}+e^-+\bar{\nu}_i$, the emitted electrons can have various energies. An emitted electron has the maximum possible energy when the electron is emitted anti-parallel to both the helium-3 nucleus and the neutrino. Then, the maximum electron energy called the endpoint energy is given by
\begin{align}
E_{\rm end}\simeq K_{\rm end}^0+m_e-m_{\rm lightest},
\label{Eend}
\end{align}
where $m_{\rm lightest}$ is the mass of the lightest neutrino. We need to distinguish the background around this endpoint from the emitted electron spectrum from the C$\nu$B with the energy, $E^{{\rm C\nu B},i}_e\simeq K_{\rm end}^0+m_e+E_{\nu_i}$ (see Eq.~(\ref{EcnuB})), since this spectrum of the inverse $\beta$-decay contains larger energies than this endpoint, $E_e^{{\rm C\nu B},i}-E_{\rm end}\simeq m_{\rm lightest}+E_{\nu_i}$. In order to estimate the rate of the background, we consider the $\beta$-decay spectrum near the endpoint \cite{Masood:2007rc}
\begin{align}
\frac{d\Gamma_\beta}{dE_e} = \frac{\bar{\sigma}}{\pi^2}N_T\sum_{i=1}^3|U_{ei}|^2H(E_e,m_{\nu_i}),
\end{align}
where $\bar{\sigma}$ is the average cross section at the leading order for neutrino capture, which is given by
\begin{align}
\bar{\sigma}=\frac{G_F^2}{2\pi}|V_{ud}|^2\frac{m_{\mathrm{^3He}}}{m_{\mathrm{^3H}}}\left(\langle f_F \rangle^2+\frac{g_A^2}{g_V^2}\langle g_{GT} \rangle^2 \right)F(2, E_e)E_e|\bm{p}_e|.
\end{align}
$H(E_e,m_{\nu_i})$ takes the following form,
\begin{align}
H(E_e, m_{\nu_i}) = \frac{1-m_e^2/(E_em_{\mathrm{^3H}})}{(1-2E_e/m_{\mathrm{^3H}}+m_e^2/m_{\mathrm{^3H}}^2)^2}
\sqrt{y_i\left(y_i+\frac{2m_{\nu_i}m_{\mathrm{^3He}}}{m_{\mathrm{^3H}}}\right)} \left[y_i+\frac{m_{\nu_i}}{m_{\mathrm{^3H}}}(m_{\mathrm{^3He}}+m_{\nu_i})\right],
\end{align}
with $y_i \simeq K_{\rm end}^0+m_e-m_{\nu_j}-E_e$.

In order to take into account the energy resolution of the detector $\Delta$, 
we model the measured spectrum as a Gaussian-smeared version of the actual spectrum. This is achieved by a convolution of both the inverse $\beta$-decay spectrum and the $\beta$-decay spectrum with a Gaussian with a full width at half maximum (FWHM) equal to $\Delta$. 
The Gaussian-smeared versions of the neutrino capture event rate and $\beta$-decay event rate are given by, respectively,
\begin{align}
\frac{d\tilde{\Gamma}_i}{dE_e} &=\frac{1}{\sqrt{2\pi}\sigma}\int^{\infty}_{-\infty}dE_e'\  \Gamma_i(E_e')\ \delta[E_e'-(E_{\rm end}+E_{\nu_i}+m_{\rm lightest})]\exp\left[-\frac{(E_e'-E_e)^2}{2\sigma^2} \right], 
\label{tildeGi}
 \\
\frac{d\tilde{\Gamma}_{\beta}}{dE_e}&=\frac{1}{\sqrt{2\pi}\sigma}\int^{\infty}_{-\infty}dE_e' \ \frac{d \Gamma_{\beta}}{dE_e}(E_e')\ \exp\left[-\frac{(E_e'-E_e)^2}{2\sigma^2} \right],
\end{align}
where $\sigma=\Delta/\sqrt{8\ln 2}$ is a standard deviation, not a cross section. After substituting Eq.~(\ref{Gi}) into Eq.~(\ref{tildeGi}), the smeared neutrino capture rate can be written as
\begin{align}
&\frac{d\tilde{\Gamma}_i}{dE_e} \nonumber \\
&=\frac{N_T}{\sqrt{2\pi}\sigma}\sum_{{s_{\nu}}=\pm\frac{1}{2}} \int \frac{d^3 p}{(2\pi)^3}\sigma_i(\bm{p}, s_{\nu})v_{\nu_i}f_{\nu_i}(\bm{p},s_{\nu})\exp \left\{-\frac{[E_e-(E_{\rm end}+m_{\rm lightest}+E_{\nu_i})]^2}{2\sigma^2} \right\},
\label{Fredholm}
\end{align}
where
\begin{align}
\sigma_i(\bm{p},s_{\nu})&=\sigma_i(\bm{p},s_{\nu},E_e') \nonumber \\
&=\sigma_i(\bm{p},s_{\nu},E_{\rm end}+m_{\rm lightest}+E_{\nu_i}).
\end{align}
Since Eq.~(\ref{Fredholm}) is a Fredholm integral equation of the first kind and $\frac{d\tilde{\Gamma}_i}{dE_e}$ is a would-be observed quantity, after solving Eq.~(\ref{Fredholm}) inversely, the spectrum of a cosmic neutrino background, $f_{\nu_i}(\bm{p},s_\nu)$, can be in principle reconstructed.

In order to discuss the potential to distinguish the C$\nu$B signal from the $\beta$-decay event, we estimate the ratio between the C$\nu$B signal and the $\beta$-decay event as done in \cite{Long:2014zva}. We define the C$\nu$B signal and $\beta$-decay event rates within a measured energy bin of width $\Delta$ centered at $E^{{\rm C\nu B},i}_e\simeq E_{\rm end}+m_{\rm lightest}+E_{\nu_i}$ as
\begin{align}
\tilde{\Gamma}_i(\Delta)=\int^{E^{{\rm C\nu B},i}_e+\Delta/2}_{E^{{\rm C\nu B},i}_e-\Delta/2}dE_e\frac{d\tilde{\Gamma}_i}{dE_e}(E_e), 
\label{OGi} \\
\tilde{\Gamma}_{\beta, i}(\Delta)=\int^{\langle E^{{\rm C\nu B},i}_e \rangle+\Delta/2}_{\langle E^{{\rm C\nu B},i}_e \rangle-\Delta/2}dE_e\frac{d\tilde{\Gamma}_\beta}{dE_e}(E_e).
\label{OGbeta}
\end{align}
Since cosmic neutrinos have various momenta, we define $\langle E^{{\rm C\nu B},i}_e \rangle$ as
\begin{align}
\langle E^{{\rm C\nu B},i}_e \rangle = E_{\rm end}+m_{\rm lightest}+ \sqrt{\langle p_0 \rangle^2 + m_{\nu_i}^2}.
\end{align}
Then the ratio between Eqs.~(\ref{OGi}) and (\ref{OGbeta}) is defined by
\begin{align}
r_{\rm C\nu B}^i(\Delta)=\frac{\tilde{\Gamma}_i(\Delta)}{\tilde{\Gamma}_{\beta, i}(\Delta)}.
\end{align}
To probe cosmological effects, such as neutrino spectral distortions from the neutrino decoupling and neutrino clustering in our Galaxy, we would have to observe C$\nu$B signals with $1\%$ precision.
Precise detection of a C$\nu$B signal within a $1\%$ precision would be successful if $r_{\rm C\nu B}^i \gg 100$ and be impossible if $r_{\rm C\nu B}^i \ll 100$. However, this estimation is almost the same as whether $r^i_{\rm C\nu B} \gg 1$ or not, because $r_{\rm C\nu B}^i$ is an exponentially rising function of $\Delta$.

Unfortunately, the precise calculation of the Gaussian smeared neutrino capture rate, Eq.~(\ref{Fredholm}), requires the knowledge of the neutrino spectral distortion from the neutrino decoupling, $\delta f_{\nu_i}^d$, and the effect of the gravitational clustering on the spectrum, $\delta f_{\nu_i}^c$, although the calculation of $\delta f_{\nu_i}^c$ is one of future work. Hereafter, we neglect these sub-leading effects, $\delta f_{\nu_i}^d$ and $\delta f_{\nu_i}^c$ since we only know the leading order of the required $\Delta$ to distinguish the signals and the background, and $\delta f_{\nu_i}^d$ and $\delta f_{\nu_i}^c$ would not affect $\Delta$ significantly.

In Tables~\ref{tb:DeltaNH} and \ref{tb:DeltaIH}, we show the required $\Delta$ to distinguish the signals and the background in Dirac and Majorana cases for $r_{\rm C\nu B}^i=1$ and $r_{\rm C\nu B}^i=100$, considering the normal and inverted mass ordering, respectively. 
In the Majorana case, we denote the ratio between the C$\nu$B signal and $\beta$-decay event as $r_{\rm C\nu B}^M$ while we denote this ratio as $r_{\rm C\nu B}^D$ in the Dirac case.
Both in the Dirac and Majorana cases, we find almost the same required $\Delta$. In the NH case, the required $\Delta$ values for $\nu_1$ with $m_{\nu_1}=0\ {\rm meV}$, $\nu_2$ with $m_{\nu_1}=8.6\ {\rm meV}$ and $\nu_3$ with $m_{\nu_1}=50\ {\rm meV}$ are $0.46\ {\rm meV}, 4.2\ {\rm meV}$ and $19\ {\rm meV}$ respectively.
In the IH case, it is difficult to distinguish the signals for $\nu_1$ and $\nu_2$ due to the degenerate masses, so we estimate the required $\Delta$ to distinguish the degenerate signals for $\nu_1$ and $\nu_2$ and the $\beta$-decay background, using the following ratio,
\begin{align} 
r_{\rm C\nu B}=r_{\rm C\nu B}^1+r_{\rm C\nu B}^2\ \ \ \ {\rm for\ \  \nu_1\ and\ \nu_2\ \ in\ the\ IH\ case}.
\end{align}
The required $\Delta$ values for $\nu_1$ and $\nu_2$ with $m_{\nu_1}=49.3\ {\rm meV}$ and $m_{\nu_2}=50\ {\rm meV}$, and $\nu_3$ with $m_{\nu_3}=0\ {\rm meV}$ are $23\ {\rm meV}$ and $0.46\ {\rm meV}$ respectively. 
If the lightest neutrino is massive, the required tiny energy resolution for the lightest neutrino becomes larger and the difficulty more relaxed.
Although the capture rate for massive Majorana neutrinos is approximately twice the rate for massive Dirac neutrinos, the required $\Delta$ values in both cases are the same, which implies that the neutrino spectral distortions from the decoupling and the neutrino clustering would not affect these $\Delta$s. 

In order to estimate the required $\Delta$ to distinguish the two degenerate cosmic neutrinos in the IH case, we estimate the following ratio,
\begin{align}
r_{12}(\Delta)&=\frac{\int^{E^{{\rm C\nu B},1}_e+\Delta/2}_{E^{{\rm C\nu B},1}_e-\Delta/2}dE_e\frac{d\tilde{\Gamma}_1}{dE_e}(E_e)}{\int^{E^{{\rm C\nu B},1}_e+\Delta/2}_{E^{{\rm C\nu B},1}_e-\Delta/2}dE_e\frac{d\tilde{\Gamma}_2}{dE_e}(E_e)}, \label{r12} \\
r_{21}(\Delta)&=\frac{\int^{E^{{\rm C\nu B},2}_e+\Delta/2}_{E^{{\rm C\nu B},2}_e-\Delta/2}dE_e\frac{d\tilde{\Gamma}_2}{dE_e}(E_e)}{\int^{E^{{\rm C\nu B},2}_e+\Delta/2}_{E^{{\rm C\nu B},2}_e-\Delta/2}dE_e\frac{d\tilde{\Gamma}_1}{dE_e}(E_e)}.
\label{r21}
\end{align}
Eq.~(\ref{r12}) characterizes the distinguishability of $\nu_1$ from $\nu_2$ from whereas Eq.~(\ref{r21}) characterizes that of $\nu_2$ from $\nu_1$. Using this, we find both in the Dirac and Majorana cases
\begin{align}
\Delta &\simeq 0.50\ {\rm meV}\ \ \ \ {\rm for}\ \  r_{12}=100, \nonumber \\
\Delta &\simeq 0.42\ {\rm meV}\ \ \ \ {\rm for}\ \  r_{21}=100. 
\end{align} 
With this energy resolution, we would be able to distinguish the signals for $\nu_1$ and $\nu_2$ in the inverted hierarchy.

\begin{figure}[htbp]
   \begin{center}
     \includegraphics[clip,width=10.0cm]{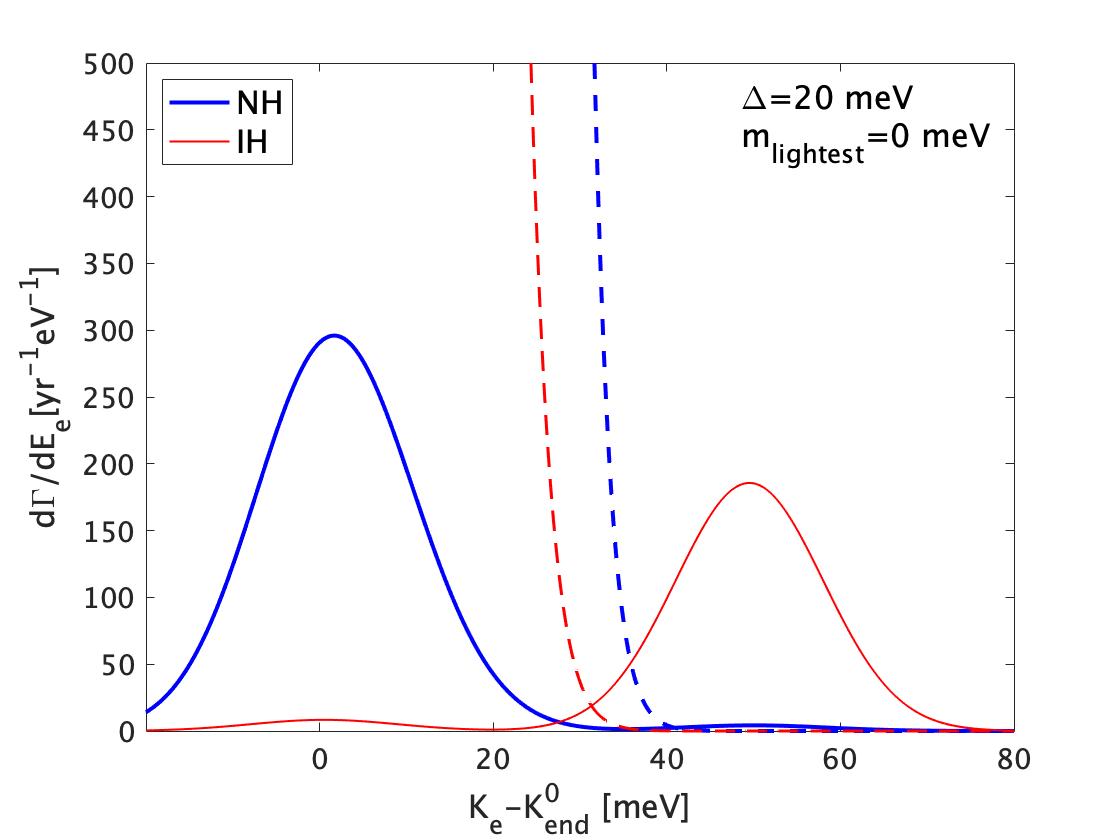}
    \end{center}
     \vspace{-3mm}
 \caption{The expected spectra of the electron kinetic energy, $K_e=E_e-m_e$, for the C$\nu$B signals (solid lines) and the $\beta$-decay background (dashed lines) in a tritium experiment, assuming 100 g of tritium, with the energy resolution $\Delta=20\ {\rm meV}$ in the case of Dirac neutrinos.
Bold blue lines represent the NH case and fine red lines represent the IH case.
We set $m_{\rm lightest}=0\ {\rm meV}$ and neglect the neutrino spectral distortions from the decoupling and the gravitational clustering effects.}
 \label{fig:SpectrumDirac20}
 \end{figure}

\begin{figure}
	\begin{center}
	\includegraphics[clip,width=10.0cm]{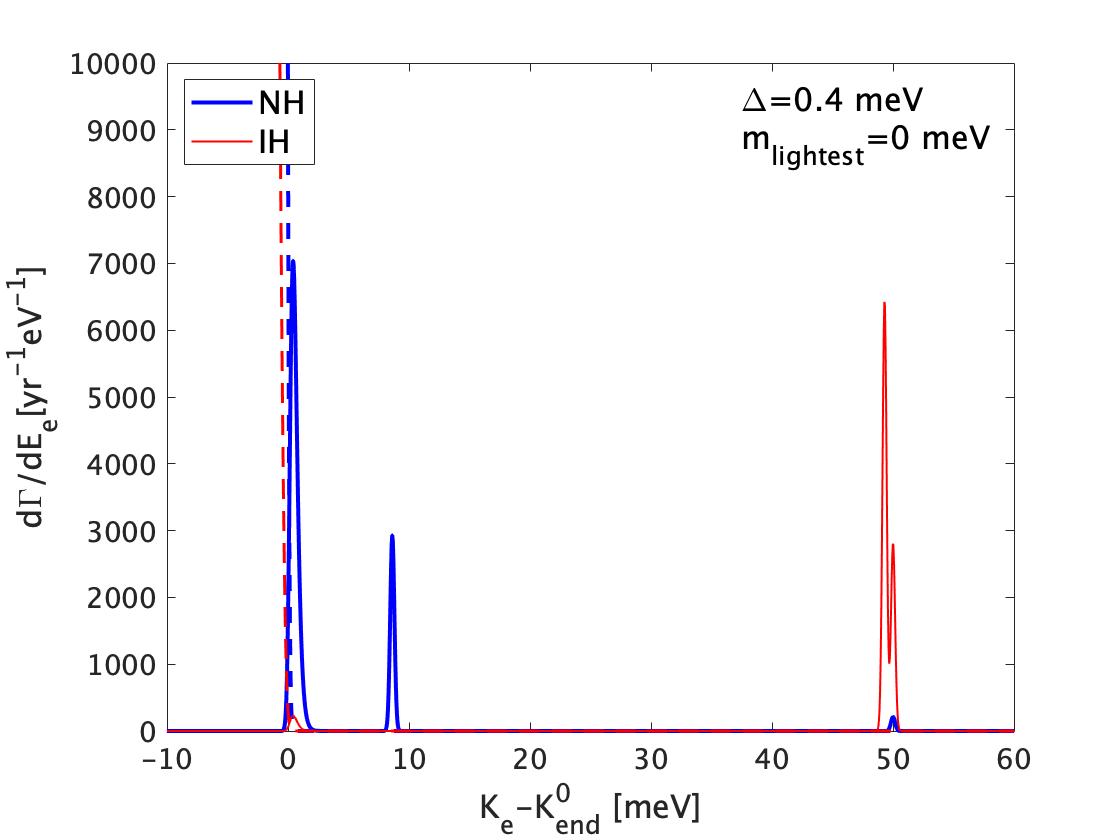}
	\end{center}
	 \vspace{-3mm}
	\caption{The expected spectra for the C$\nu$B signals (solid lines) and the $\beta$-decay background (dashed lines) in a tritium experiment, assuming 100 g of tritium, with the energy resolution $\Delta=0.4\ {\rm meV}$ in the case of Dirac neutrinos. We also set $m_{\rm lightest}=0\ {\rm meV}$ and neglect the neutrino spectral distortions from the decoupling and the gravitational clustering effects.}
	\label{fig:SpectrumDirac04}
\end{figure}

In fact, we can see that the C$\nu$B signals and the background would be separated with these values of $\Delta$.
In Figs~\ref{fig:SpectrumDirac20} and \ref{fig:SpectrumDirac04}, we show the expected C$\nu$B spectra and background of the electron kinetic energy, $K_e=E_e-m_e$, for $\Delta=20\ {\rm meV}$ and $0.4\ {\rm meV}$ respectively, considering Dirac neutrinos and both hierarchies. In these figures, we also neglect the neutrino spectral distortions from the decoupling and the gravitational clustering effects. The characteristic peaks of the C$\nu$B signals exist at $K_e^{{\rm C\nu B},i}-K_{\rm end}^0 \simeq E_{\nu_i}$.

For the case of $\Delta=20\ {\rm meV}$, the capture events for the heaviest neutrinos can be resolved from the $\beta$-decay background in both hierarchies. The $\beta$-decay background near the end point, $E_{\rm end}$, in the NH case is larger than that in the IH case since the lightest neutrinos contribute to the endpoint most efficiently through $|U_{e i}|^2$, and $|U_{e i}|^2$ for the lightest neutrinos in the NH case is larger than that of the IH case.

For the extremely small energy resolution of $\Delta =0.4\ {\rm meV}$, we can completely distinguish the spectra of all three mass eigenstates of cosmic neutrinos and the background. In addition, in the IH case, the degenerate spectra of $\nu_1$ and $\nu_2$ can also be resolved when $\Delta=0.4\ {\rm meV}$.
From these results, we can conclude that the bound on the required $\Delta$ is $\Delta=0.4\ {\rm meV}$ in a neutrino capture experiment on tritium.

\begin{table}[h]
\begin{center}
	\begin{tabular}{|l|l|l|l|l|l|l|}
		\hline
		 NH case& $r_{\rm C\nu B}^M=1$ & $r_{\rm C\nu B}^M=100$ & $r_{\rm C\nu B}^D=1$ & $r_{\rm C\nu B}^D=100$ \\
		\hline
		$\nu_1\ (m_{\nu_1}=0\ {\rm meV})$ & $\Delta =0.83\ {\rm meV}$ &  $\Delta =0.46\ {\rm meV}$ & $\Delta =0.83\ {\rm meV}$ & $\Delta =0.46\ {\rm meV}$   \\
		$\nu_2\ (m_{\nu_2}=8.6\ {\rm meV})$ & $\Delta =5.3\ {\rm meV}$ &  $\Delta =4.2\ {\rm meV}$ & $\Delta =5.1\ {\rm meV}$ & $\Delta =4.1\ {\rm meV}$   \\
		$\nu_3\ (m_{\nu_3}=50\ {\rm meV})$ & $\Delta =21\ {\rm meV}$ &  $\Delta =19\ {\rm meV}$ & $\Delta =21\ {\rm meV}$ & $\Delta =18\ {\rm meV}$   \\
		\hline
	\end{tabular}
	\caption{The required $\Delta$ to distinguish the C$\nu$B signal and the $\beta$-decay background for the various $\nu_i$ with the mass $m_{\nu_i}$, $r_{\rm C\nu B}$ and Dirac and Majorana cases in the normal mass ordering. Here we neglect the neutrino spectral distortions from the neutrino decoupling, $\delta f_{\nu_i}^d$, and the effect of the gravitational clustering on the spectrum, $\delta f_{\nu_i}^c$, since we only estimate the leading order of the required $\Delta$.}
  \label{tb:DeltaNH}
\end{center}
\end{table}

\begin{table}[h]
\begin{center}
	\begin{tabular}{|l|l|l|l|l|l|l|}
		\hline
		 IH case& $r_{\rm C\nu B}^M=1$ & $r_{\rm C\nu B}^M=100$ & $r_{\rm C\nu B}^D=1$ & $r_{\rm C\nu B}^D=100$ \\
		\hline
		$\nu_1+\nu_2 $ & $\Delta =29\ {\rm meV}$ &  $\Delta =23\ {\rm meV}$ & $\Delta =28\ {\rm meV}$ & $\Delta =22\ {\rm meV}$   \\
		\hspace{-0.1cm}$(m_{\nu_1, \nu_2}=49.3, 50\ {\rm meV})$ \hspace{-0.3cm} & &   &  &    \\
		$\nu_3\ (m_{\nu_3}=0\ {\rm meV})$ & $\Delta =0.83\ {\rm meV}$ &  $\Delta =0.46\ {\rm meV}$ & $\Delta =0.83\ {\rm meV}$ & $\Delta =0.46\ {\rm meV}$   \\
		\hline
	\end{tabular}
	\caption{The required $\Delta$ to distinguish  the C$\nu$B signal and the $\beta$-decay background for the several $\nu_i$ with the mass $m_{\nu_i}$, $r_{\rm C\nu B}$ and Dirac and Majorana cases in the inverted mass ordering. Here we also neglect $\delta f_{\nu_i}^d$ and $\delta f_{\nu_i}^c$ since we only estimate the leading order of the required $\Delta$.}
  \label{tb:DeltaIH}
\end{center}
\end{table}

\section{Summary and discussion}
\label{sec:5}

We have discussed precise estimates of the expected capture rate of neutrinos from the C$\nu$B on a tritium target, including cosmological effects such as neutrino spectral distortions from the neutrino decoupling and the gravitational clustering in our Galaxy. After formulating such a capture rate, we have concretely computed the capture rates of each neutrino species for both Dirac and Majorana neutrinos in both the normal and the inverted hierarchies of the neutrino mass spectrum.

These precise estimates of the capture rates are important in two ways. Firstly, this precise calculation, once matched with observation, will allow for a probe into the early universe. That is, the detailed process of the neutrino decoupling as well as the dark matter distribution through the clustering effects of non-relativistic neutrinos will be illuminated. Secondly, once one would be able to find the deviation from this precise estimate of the capture rate, the possible deviation might suggest new
neutrino physics and/or non-standard evolution of the universe. 

Such cosmological effects modify the neutrino capture rates mainly through the neutrino number densities.
In order to estimate the impact on the capture rates, we have computed the precise number densities of neutrinos in the current universe.
The neutrino spectral distortions from the neutrino decoupling change the number densities by $1.1\%$ for $\nu_1$, $1.0\%$ for $\nu_2$ and $0.9\%$ for $\nu_3$ whereas the gravitational clustering effects modify those by $12\%$ for $m_{\nu_i}=50\ {\rm meV}$ and $0.53\%$ for $m_{\nu_i}=10\ {\rm meV}$ \cite{Zhang:2017ljh}. The estimated errors of the neutrino capture rates mainly come from the uncertainties of parameters of neutrino mixing, $|U_{ei}|^2$, and the reduced matrix element of the GT operator of tritium, $\langle g_{\rm GT} \rangle$. The current errors of the PMNS matrix are about $10\%$ at $3\sigma$ level \cite{Esteban:2020cvm}. The theoretical calculation of $\langle g_{\rm GT} \rangle$ still includes the uncertainty of a few $\%$, although the estimation of $\langle g_{\rm GT} \rangle$ through the observation of the tritium half-life and the value of the Fermi operator, $\langle f_F \rangle$, only involves an uncertainty of $0.1\%$ \cite{Baroni:2016xll}.
In order to observe such cosmological effects through cosmic neutrino capture on tritium, one needs to have about $10^4$ events since one needs to measure the signal with $1\%$ precision.
To achieve this goal, one would need about 10 kg of tritium due to the half-life of tritium of 12.32 years, and the uncertainties of the PMNS matrix and the reduced matrix element of GT operator must be improved to within $0.1\%$ level in future.
Planned neutrino oscillation experiments are expected to improve the precision of the PMNS matrix in the near future (see e.g. \cite{Abe:2016tii, Abe:2018uyc, Abi:2020evt}).
In addition, to distinguish between the gravitational clustering effect and the spectral distortions from the neutrino decoupling for massive neutrinos, we will need to improve the computation of gravitational clustering and spectral distortions from the neutrino decoupling on neutrino number densities.
Since the lightest cosmic neutrinos in the Standard Model are expected not to cluster significantly in our Galaxy while massive neutrinos are, the lightest ones can contain a wealth of clean information about the physics in the early universe. To this end, much better energy resolution is required than is currently attainable.

We have also comprehensively discussed the required energy resolutions of $\Delta$ to detect each neutrino species as well as the reconstruction of the spectrum of a cosmic neutrino background.
To distinguish the C$\nu$B signals for massless neutrinos from the $\beta$-decay background, we find the required $\Delta$ is $\Delta\simeq 0.4\ {\rm meV}$. In addition, to resolve the C$\nu$B signals for degenerate $\nu_1$ and $\nu_2$ in the IH case, we also identify the required $\Delta$ as $\Delta \simeq 0.4\ {\rm meV}$. With this energy resolution, one can completely resolve the signals for the mass-eigenstates of neutrinos and the background.


\section*{Acknowledgments}

KA is supported by JSPS Grant-in-Aid for Research Fellows
No. 19J14449. KA and MY are supported in part by JSPS Bilateral Open
Partnership Joint Research Projects. SH is supported by the Japanese Government MEXT Scholarship. MY is supported in part by JSPS
Grant-in-Aid for Scientific Research Numbers 18K18764 and Mitsubishi
Foundation.


\appendix

\section{Exact neutrino capture rate at tree level}
\label{appa}
In this appendix, we consider the exact neutrino capture rate at tree level on a tritium target for the first time.
This exact capture rate at tree level may be useful for estimating the value of the capture rate including tiny cosmological effect such as the anisotropy of the C$\nu$B in the early universe.
To begin, following the path of \cite{Long:2014zva}, we calculate the polarized neutrino capture cross section of inverse beta decay of a single neutron: $\nu + n \rightarrow p + e^-$.
The matrix element in the cross section comes from the 4-Fermi Lagrangian, which is applicable here as the energies of the particles are all much lower than the weak scale. In this case, the matrix element for each species is given by 
\begin{align}
i\mathcal{M}_i = -i\frac{G_F}{\sqrt{2}}V_{ud}U^*_{ei}\left[\bar{u}_e\gamma^\alpha(1-\gamma^5)u_{\nu_i}\right]\left[\bar{u}_p\gamma_{\alpha}(f-g\gamma^5)u_n\right].
\end{align}
In the above, the $f=f(q=0)$ and $g=g(q=0)$ constants are nucleonic form factors for the proton and neutron in the limit of small momentum transition.
Thus, the squared matrix element becomes 
\begin{align}
|\mathcal{M}_i|^2=\frac{G_F^2}{2}|V_{ud}|^2|U_{ei}|^2\mathcal{N}_1^{\alpha\beta}\mathcal{N}_{2\alpha\beta}
\end{align}
wherein we have
\begin{align}
\mathcal{N}_1^{\alpha \beta} &= {\rm tr}\left[\gamma^{\alpha}(1-\gamma^5)u_{\nu}\bar{u}_{\nu}\gamma^{\beta}(1-\gamma^5)u_e\bar{u}_e \right],  \\
\mathcal{N}_2^{\gamma \delta} &= {\rm tr}\left[\gamma^{\gamma}(f-g\gamma^5)u_{n}\bar{u}_{n}\gamma^{\delta}(f-g\gamma^5)u_p\bar{u}_p \right].
\end{align}
After summing the spins of the outgoing electron and proton and averaging the spins of incoming neutron, the squared matrix element is given by
\begin{align}
\frac{1}{2}\sum_{s_n,s_e,s_p=\pm 1/2}|\mathcal{M}_i|^2=
\frac{G_F^2}{4}|V_{ud}|^2|U_{ei}|^2\tilde{\mathcal{N}}_1^{\alpha\beta}\tilde{\mathcal{N}}_{2\alpha\beta},
\end{align}
where
\begin{align}
\tilde{\mathcal{N}}_1^{\alpha \beta} &= \sum_{s_e=\pm 1/2}{\rm tr}\left[\gamma^{\alpha}(1-\gamma^5)u_{\nu_i}\bar{u}_{\nu_i}\gamma^{\beta}(1-\gamma^5)u_e\bar{u}_e \right], \\
\tilde{\mathcal{N}}_2^{\gamma \delta} &= \sum_{s_n,s_p=\pm 1/2}{\rm tr}\left[\gamma^{\gamma}(f-g\gamma^5)u_{n}\bar{u}_{n}\gamma^{\delta}(f-g\gamma^5)u_p\bar{u}_p \right].
\label{mathN}
\end{align}
Using the completeness relations, we get the relation of a Dirac spinor for a neutron, a proton, and an electron,
\begin{align}
\sum_{s_j=\pm 1/2}u_j\bar{u}_j=(\slashed{p}_j+m_j),
\end{align}
and for neutrinos,
\begin{align}
u_{\nu_i}\overline{u}_{\nu_i}=\frac{1}{2}\bigl(\slashed{p}_{\nu_i}+m_{\nu_i}\bigl)\bigl(1+2s_{\nu}\gamma^5\slashed{S}_{\nu_i}\bigl)
\end{align}
where $S_{\nu_i}$ is the spin vector of a neutrino given by
\begin{align}
(S_{\nu_i})^{\alpha}=\left(\frac{|\bm{p}_{\nu}|}{m_{\nu_i}}, \frac{E_{\nu}}{m_{\nu_i}}\frac{\bm{p}_{\nu}}{|\bm{p}_\nu|}\right).
\end{align}
Note that in the massless limit, the previous relation of the Dirac spinor for neutrinos becomes
\begin{align}
u_{\nu_i}\overline{u}_{\nu_i}=\frac{1}{2}\slashed{p}_{\nu_i}\left(1-2s\gamma^5 \right),
\end{align}
where we used $mS^{\mu}=p^{\mu}$ and $p_{\mu}S^{\mu}=0$.
Using the above relations, we get Eq.~(\ref{mathN}) as
\begin{align}
\tilde{\mathcal{N}}_1^{\alpha\beta}&=\frac{1}{2}\mathrm{tr}\left[\gamma^\alpha\bigl(1-\gamma^5\bigl)\bigl(\slashed{p}_{\nu_i}+m_{\nu_i}\bigl)\bigl(1+2s_\nu\gamma^5\slashed{S}_{\nu_i}\bigl)\gamma^\beta\bigl(1-\gamma^5\bigl)\bigl(\slashed{p}_e+m_e\bigl)\right], 
\label{N2}  \\
\tilde{\mathcal{N}}_2^{\gamma\delta}&=\mathrm{tr}\left[\gamma^\gamma\bigl(f-g\gamma^5\bigl)\bigl(\slashed{p}_n+m_n\bigl)\gamma^\delta\bigl(f-g\gamma^5\bigl)\bigl(\slashed{p}_p+m_p\bigl)\right].
\label{N3}
\end{align}
Putting Eqs.~(\ref{N2}) and (\ref{N3}) together, we have
\begin{align}
&\tilde{\mathcal{N}}_1^{\alpha\beta}\tilde{\mathcal{N}}_{2\alpha\beta}= \nonumber \\
&32\left\{\left(g+f\right)^2\left[\left(p_e\cdot p_p\right)\left(p_{\nu_i} \cdot p_n\right)\right]+\left(g-f\right)^2\left[\left(p_e\cdot p_n\right)\left(p_{\nu_i}\cdot p_p\right)\right]+\left(g^2-f^2\right)m_nm_p\left(p_e\cdot p_{\nu_i}\right)\right\} \nonumber \\
&-64s_\nu m_{\nu_i} \left\{\left(g+f\right)^2\left[\left(p_e\cdot p_p\right)\left(S_{\nu_i} \cdot p_n\right)\right]+\left(g-f\right)^2\left[\left(p_e\cdot p_n\right)\left(S_{\nu_i}\cdot p_p\right)\right]+\left(g^2-f^2\right)m_nm_p\left(p_e\cdot S_{\nu_i}\right)\right\}
\end{align}

In the following, we consider the rest frame of the neutron where
\begin{align}
(p_n)^\alpha = (m_n, \bm{0}),\ \ \ \ (p_{\nu})^\alpha=(E_{\nu}, \bm{p}_{\nu}),\ \ \ \ (p_p)^{\alpha}=(E_p, \bm{p}_p),\ \ \ \ (p_e)^{\alpha}=(E_e, \bm{p}_e).
\end{align}
In this frame, we obtain
\begin{align}
\tilde{\mathcal{N}}_1^{\alpha\beta}\tilde{\mathcal{N}}_{2\alpha\beta}&=32m_nE_pE_eE_{\nu_i}\biggl\{\left(g+f\right)^2\left(1-\frac{\bm{p}_e\cdot\bm{p}_p}{E_eE_{p}}\right)+(g-f)^2\left(1-\frac{\bm{p}_{\nu}\cdot \bm{p}_p}{E_{\nu_i}E_p} \right) \nonumber \\
&\ \ \ \ +(g^2-f^2)\frac{m_p}{E_p}\left(1-\frac{\bm{p}_e\cdot\bm{p}_{\nu}}{E_eE_{\nu_i}}\right) \biggl\} -64s_{\nu}m_nE_pE_eE_{\nu_i}\biggl\{v_{\nu_i}(g+f)^2\left(1-\frac{\bm{p}_e\cdot\bm{p}_p}{E_eE_p}\right) \nonumber \\
&\ \ \ \ +(g-f)^2\left(v_{\nu_i}-\frac{\bm{p}_{\nu}\cdot\bm{p}_p}{|\bm{p}_{\nu}|E_p}\right) +(g^2-f^2)\frac{m_p}{E_p}\left(v_{\nu_i}-\frac{\bm{p}_{\nu}\cdot\bm{p}_e}{|\bm{p}_{\nu}|E_e} \right) \biggl\} \nonumber \\
&=32m_nE_pE_eE_{\nu_i}\biggl\{2\left(g^2+f^2 \right)\left(1-2s_{\nu}v_{\nu_i}\right)
+\left(g^2-f^2 \right)\frac{m_p}{E_p}\left(1-2s_{\nu}v_{\nu_i}\right) \nonumber \\
&\ \ \ \ +\left(f^2-g^2\right)\frac{m_p}{E_p}\left(v_{\nu_i}-2s_{\nu}\right)v_e\cos\theta_{e\nu} 
-\left(g+f\right)^2(1-2s_{\nu}v_{\nu_i})v_ev_p\cos\theta_{ep} \nonumber \\
&\ \ \ \ -\left(g-f\right)^2(v_{\nu_i}-2s_{\nu})v_p\cos\theta_{\nu p} \biggl\},
\label{NN}
\end{align}
where $v_j=|\bm{p}_j|/E_j$ and $\cos\theta_{jk}=\bm{p}_j\cdot\bm{p}_k/(|\bm{p}_j||\bm{p}_k|)$. Note that we do not neglect the proton recoil here. Due to the energy-momentum conservation, we can rewrite $\cos\theta_{ep}$ and $\cos\theta_{\nu p}$ as
\begin{align}
\cos\theta_{ep}=-\frac{|\bm{p}_e|}{|\bm{p}_p|}+\frac{|\bm{p}_{\nu}|}{|\bm{p}_p|}\cos\theta_{e\nu},\ \ \ \ 
\cos\theta_{\nu p}=\frac{|\bm{p}_{\nu}|}{|\bm{p}_p|}-\frac{|\bm{p}_e|}{|\bm{p}_p|}\cos\theta_{e\nu}.
\label{EMC}
\end{align}
Substituting Eq.~(\ref{EMC}) into Eq.~(\ref{NN}), we have
\begin{align}
\tilde{\mathcal{N}}_1^{\alpha\beta}\tilde{\mathcal{N}}_{2\alpha\beta}&=32m_nE_pE_eE_{\nu_i}\biggl\{2\left(g^2+f^2 \right)\left(1-2s_{\nu}v_{\nu_i}\right)
+\left(g^2-f^2 \right)\frac{m_p}{E_p}\left(1-2s_{\nu}v_{\nu_i}\right) \nonumber \\
&\ \ \ \ +\left(f^2-g^2\right)\frac{m_p}{E_p}\left(v_{\nu_i}-2s_{\nu}\right)v_e\cos\theta_{e\nu} 
+\left(g+f\right)^2(1-2s_{\nu}v_{\nu_i})v_e\left(\frac{|\bm{p}_e|}{E_p}-\frac{|\bm{p}_{\nu}|}{E_p}\cos\theta_{e\nu} \right) \nonumber \\
&\ \ \ \  -\left(g-f\right)^2(v_{\nu_i}-2s_{\nu})\biggl(\frac{|\bm{p}_{\nu}|}{E_p}-\frac{|\bm{p}_e|}{E_p}\cos\theta_{e\nu} \biggl) \biggl\}.
\end{align}
The squared matrix amplitude is given by
\begin{align}
&\frac{1}{2}\sum_{s_n,s_e,s_p=\pm 1/2}|\mathcal{M}_i|^2= \nonumber \\
&\ \ \ \ 8G_F^2|V_{ud}|^2|U_{ei}|^2m_nE_pE_eE_{\nu_i} \biggl\{2\left(g^2+f^2 \right)\left(1-2s_{\nu}v_{\nu_i}\right)
+\left(g^2-f^2 \right)\frac{m_p}{E_p}\left(1-2s_{\nu}v_{\nu_i}\right) \nonumber \\
&\ \ \ \ +\left(f^2-g^2\right)\frac{m_p}{E_p}\left(v_{\nu_i}-2s_{\nu}\right)v_e\cos\theta_{e\nu} 
+\left(g+f\right)^2(1-2s_{\nu}v_{\nu_i})v_e\left(\frac{|\bm{p}_e|}{E_p}-\frac{|\bm{p}_{\nu}|}{E_p}\cos\theta_{e\nu} \right) \nonumber \\
&\ \ \ \  -\left(g-f\right)^2(v_{\nu_i}-2s_{\nu})\biggl(\frac{|\bm{p}_{\nu}|}{E_p}-\frac{|\bm{p}_e|}{E_p}\cos\theta_{e\nu} \biggl) \biggl\}.
\label{M2}
\end{align}

In the center-of-mass frame, the differential cross section is
\begin{align}
\frac{d\sigma_i}{dt}=\frac{1}{64 \pi s}\frac{1}{|\bm{p}_{\nu}^{\rm cm}|^2}\ \frac{1}{2}\sum_{s_n,s_e,s_p=\pm 1/2}|\mathcal{M}_i|^2,
\end{align}
where $s=(p_n+p_{\nu_i})^2$ and $d\Omega=d\cos\theta_{e\nu}d\psi$. $\bm{p}^{\rm cm}$ denotes a momentum in the center-of-mass frame. $t$ and $\bm{p}_{\nu}^{\rm cm}$ can be expressed in the rest frame of neutron as
\begin{align}
t&= (E_e-E_{\nu_i})^2-|\bm{p_e}-\bm{p}_{\nu}|^2=(E_e-E_{\nu_i})^2-|\bm{p}_e|^2-|\bm{p}_{\nu}^2|+2|\bm{p}_e||\bm{p}_{\nu}|\cos\theta_{e\nu_i}, \nonumber \\
\bm{p}_{\nu}^{\rm cm}&=\bm{p}_{\nu}\frac{m_n}{\sqrt{s}}.
\end{align}
Then the differential cross section in the rest frame of the neutron is given by
\begin{align}
\frac{d \sigma_i}{d \cos\theta_{e\nu_i}}=\frac{1}{32\pi}\frac{1}{m_n^2}\frac{|\bm{p}_e|}{|\bm{p}_{\nu}|}\  \frac{1}{2}\sum_{s_n,s_e,s_p=\pm 1/2}|\mathcal{M}_i|^2.
\label{DCS}
\end{align}
Using Eq.~(\ref{M2}) in Eq.~(\ref{DCS}) and including the Fermi function, we obtain
\begin{align}
\frac{d \sigma_i}{d \cos\theta_{e\nu_i}}&=\frac{G_F^2}{4\pi}|V_{ud}|^2|U_{ei}|^2F(Z,E_e)\frac{E_pE_e|\bm{p}_e|}{m_nv_{\nu_i}} \nonumber \\
&\ \ \ \ \times \biggl\{2\left(g^2+f^2 \right)\left(1-2s_{\nu}v_{\nu_i}\right)
+\left(g^2-f^2 \right)\frac{m_p}{E_p}\left(1-2s_{\nu}v_{\nu_i}\right) \nonumber \\
&\ \ \ \ +\left(f^2-g^2\right)\frac{m_p}{E_p}\left(v_{\nu_i}-2s_{\nu}\right)v_e\cos\theta_{e\nu} 
+\left(g+f\right)^2(1-2s_{\nu}v_{\nu_i})v_e\left(\frac{|\bm{p}_e|}{E_p}-\frac{|\bm{p}_{\nu}|}{E_p}\cos\theta_{e\nu} \right) \nonumber \\
&\ \ \ \  -\left(g-f\right)^2(v_{\nu_i}-2s_{\nu})\biggl(\frac{|\bm{p}_{\nu}|}{E_p}-\frac{|\bm{p}_e|}{E_p}\cos\theta_{e\nu} \biggl) \biggl\}.
\end{align}
After integrating over $\theta_{e\nu_i}$, the total capture cross section is
\begin{align}
\sigma_i&=\frac{G_F^2}{2\pi}|V_{ud}|^2|U_{ei}|^2F(Z,E_e)\frac{E_pE_e|\bm{p}_e|}{m_nv_{\nu_i}} \nonumber \\
&\ \ \ \ \times \biggl\{2\left(g^2+f^2 \right)\left(1-2s_{\nu}v_{\nu_i}\right)
+\left(g^2-f^2 \right)\frac{m_p}{E_p}\left(1-2s_{\nu}v_{\nu_i}\right) \nonumber \\
&\ \ \ \  +\left(g+f\right)^2(1-2s_{\nu}v_{\nu_i})v_e\frac{|\bm{p}_e|}{E_p}  -\left(g-f\right)^2(v_{\nu_i}-2s_{\nu})\frac{|\bm{p}_{\nu}|}{E_p} \biggl\}.
\label{CS1}
\end{align}
Note that we neglect an angular dependence of $E_p$ for simplicity since the angular dependence is extremely small as
\begin{align}
E_p &\simeq m_p\left(1+\frac{|\bm{p}_p|}{2m_p^2}\right), \nonumber \\
&\simeq m_p\left(1+ \frac{|\bm{p}_e|^2}{2m_p^2}-\frac{|\bm{p}_e||\bm{p}_{\nu}|}{m_p^2}\cos\theta_{e\nu_i} \right),
\label{ExactEp}
\end{align}
where $|\bm{p}_p|=|\bm{p}_e-\bm{p}_{\nu}|$. From Eq.~(\ref{ExactEp}), the angle-dependent term is suppressed by $\mathcal{O}(10^{-18})$ compared to the leading order contribution.

Finally, we comment on the magnitude of each term in Eq.~(\ref{CS1}). The leading order terms independent of $v_{\nu_i}$ come from the second line in Eq.~(\ref{CS1}) under the approximation of $E_p\simeq m_p$. The next-to-leading order contributions come from the terms proportional to $\nu_i$ in the second line in Eq.~(\ref{CS1}) under the approximation of $E_p\simeq m_p$, which depend on $m_{\nu_i}$ and are suppressed by $\mathcal{O}(1-10^{-2})$ when compared with the leading order. The next-to-next-to leading order (NNLO) terms are those proportional to $|\bm{p}_e|/E_p\simeq |\bm{p_e}|/m_p$, which are suppressed by about $\mathcal{O}(10^{-5}-10^{-7})$ when compared with the leading terms. The NNNLO terms are the terms proportional to $|\bm{p}_e|/(2m_p^2)$, which is $\mathcal{O}(10^{-9})$. The NNNNLO terms come from those proportional to $|\bm{p}_{\nu}|/E_p$, which is $\mathcal{O}(10^{-13})$. The last contribution comes from those proportional to $|\bm{p}_e||\bm{p}_{\nu}|/m_p^2$, which is $\mathcal{O}(10^{-18})$. But it should be noted that one-loop corrections to the cross section and other atomic corrections might be larger than them.

To account for tritium inverse beta decay rather than that of a neutron, the following changes are made: the neutron and proton masses become the tritium and helium-3 masses respectively, and the nucleonic form factors are replaced by transition probabilities, with $f^2$ becoming $\braket{f_F}^2$ and $3g^2$ becoming $\frac{g_A^2}{g_V^2}\braket{g_{GT}}^2$ \cite{Baroni:2016xll}.


\section{Kinematics}
\label{appb}
In this appendix, we evaluate the kinematics of tritium beta decay and inverse tritium beta decay for cosmic neutrinos. 
In particular, we investigate the maximal kinetic energy of an electron emitted from $\beta$-decay, called the $\beta$-decay endpoint kinetic energy, and the kinetic energy of an electron emitted from the inverse $\beta$-decay process for the C$\nu$B.
Here we only consider the nuclear process, although what we really should discuss is the atomic process.

We first consider the kinematics of tritium beta decay, $\mathrm{^3H}\rightarrow \mathrm{^3He}+e^-+\bar{\nu}_i$, in the rest frame of $\mathrm{^3H}$.
From 4-momentum conservation, the kinetic energy of the electron, which is defined as $K_e=E_e-m_e$, is
\begin{align}
K_e=\frac{(m_{\mathrm{^3H}}-m_e)^2-m_{\nu_i}^2-m_{\mathrm{^3He}}^2-2E_{\nu_i}E_{\mathrm{^3He}}+2|\bm{p}_{\nu}||\bm{p}_{\mathrm{^3He}}|\cos \theta_{\nu \mathrm{^3He}}}{2m_{\mathrm{^3H}}}.
\end{align}
The maximal kinetic energy, $K_{\rm end}$, is achieved when the emitted anti-neutrino is the lightest and $\cos \theta_{\nu \mathrm{^3He}}=1\ (\theta_{\nu \mathrm{^3He}}=0)$. 
When the neutrino and the helium-3 nucleus are emitted in parallel, the electron is produced in anti-parallel. 
In addition, the maximization condition of the electron energy corresponds to the minimization condition of $(E_{\nu}+E_{\mathrm{^3He}})$, which yields
\begin{align}
\frac{|\bm{p}_{\nu}|}{|\bm{p}_{\mathrm{^3He}}|}={\frac{m_{\nu_i}}{m_{\mathrm{^3He}}}}.
\end{align}
From these conditions, the maximal kinetic energy of the electron is given by
\begin{align}
K_{\rm end}=\frac{(m_{\mathrm{^3H}}-m_e)^2-(m_{\rm lightest}+m_{\mathrm{^3He}})^2}{2m_{\mathrm{^3H}}}.
\end{align}  
If the lightest neutrino is massless, the endpoint kinetic energy is identified as
\begin{align}
K_{\rm end}^0=\frac{(m_{\mathrm{^3H}}-m_e)^2-m_{\mathrm{^3He}}^2}{2m_{\mathrm{^3H}}}.
\end{align}
Under the approximation, $m_{\mathrm{^3H}}\simeq m_{\mathrm{^3He}}$, the difference between the endpoint kinetic energy for massive and massless neutrinos is
\begin{align}
\Delta K^0 &= K_{\rm end}-K_{\rm end}^0 \nonumber \\
&\simeq-m_{\rm lightest}.
\end{align} 

Next we investigate the kinematics of inverse tritium beta decay for relic neutrinos, $\nu_i + \mathrm{^3H}\rightarrow \mathrm{^3He}+e^-$. In the rest-frame of $\mathrm{^3H}$, we similarly get the kinetic energy of the electron as
\begin{align}
K_e^{\rm{C\nu B}}&=\frac{(E_{\nu_i}+m_{\mathrm{^3H}}-m_e)^2-|\bm{p}_{\nu}|^2+2|\bm{p}_{\nu}||\bm{p}_e|\cos \theta_{e\nu}-m_{\mathrm{^3He}}^2}{2(E_{\nu_i}+m_{\mathrm{^3H}})} \nonumber \\
&\simeq \frac{(E_{\nu_i}+m_{\mathrm{^3H}}-m_e)^2-m_{\mathrm{^3He}}^2}{2(E_{\nu_i}+m_{\mathrm{^3H}})}.
\end{align}
where we neglect the terms proportional to $|\bm{p}_{\nu}|^2$ and $|\bm{p}_{\nu}||\bm{p}_e|$ and leave the term proportional to $E_{\nu_i}m_{\mathrm{^3H}}$ because of $m_{\mathrm{^3H}} \gg |\bm{p}_e| \gg |\bm{p}_\nu|$.
For $m_{\mathrm{^3H}} \gg m_e$, the difference between $K_e^{\rm C\nu B}$ and $K_{\rm end}^0$ is
\begin{align}
\Delta K^{\rm C\nu B} &= K_e^{\rm C\nu B}-K_{\rm end}^0 \nonumber \\
&\simeq E_{\nu_i}.
\end{align}
Since $\Delta K^0$ and $\Delta K^{\rm C\nu B}$ are (approximately) not functions of any nuclear masses, they are insensitive to the uncertainties in the nuclear masses which are calculated from the measured values of atomic masses.


\end{document}